\documentclass[preprint2]{proto}
\usepackage{times,color}
\newcommand{\refs}{\par\noindent\hangindent=1pc\hangafter=1}
\begin{document}

\title{\textbf{\LARGE Multiplicity in Early Stellar Evolution}}

 \author {\textbf{\large Bo Reipurth$^1$,
  Cathie J. Clarke$^2$,
Alan P. Boss$^3$, 
Simon P. Goodwin$^4$,
 Luis Felipe Rodr\'\i guez$^5$,
 Keivan G. Stassun$^6$,
 Andrei Tokovinin$^7$,
 Hans Zinnecker$^8$}
\\~\\
}

\bigskip

\affil{\small\em 1: University of Hawaii,
 2: Institute of Astronomy, Cambridge,
 3: Carnegie Institution of Washington,
 4: University of Sheffield,
 5: Universidad Nacional Aut\'onoma de Mexico,
 6: Vanderbilt University,
 7: Cerro Tololo Interamerican Observatory,
 8: NASA-Ames
\\}

\begin{abstract}
\baselineskip = 11pt
\leftskip = 0.65in 
\rightskip = 0.65in
\parindent=1pc {\small Observations from optical to centimeter
  wavelengths have demonstrated that multiple systems of two or more
  bodies is the norm at all stellar evolutionary stages. Multiple
  systems are widely agreed to result from the collapse and
  fragmentation of cloud cores, despite the inhibiting influence of
  magnetic fields. Surveys of Class 0 protostars with mm
  interferometers have revealed a very high multiplicity frequency of
  about 2/3, even though there are observational difficulties in
  resolving close protobinaries, thus supporting the possibility that
  all stars could be born in multiple systems. Near-infrared adaptive
  optics observations of Class I protostars show a lower binary
  frequency relative to the Class~0 phase, a declining trend that
  continues through the Class II/III stages to the field population.
  This loss of companions is a natural consequence of dynamical
  interplay in small multiple systems, leading to ejection of members.
  We
  discuss observational consequences of this dynamical evolution, and
  its influence on circumstellar disks, and we review the evolution of
  circumbinary disks and their role in defining binary mass ratios.
  Special attention is paid to eclipsing PMS binaries, which allow for
  observational tests of evolutionary models of early stellar
  evolution. Many stars are born in clusters and small groups, and we
  discuss how interactions in dense stellar environments can
  significantly alter the distribution of binary separations through
  dissolution of wider binaries. The binaries and multiples we find in
  the field are the survivors of these internal and external
  destructive processes, 
  and we
  provide a detailed overview of the multiplicity statistics of the
  field, which form a boundary condition for all models of binary
  evolution. Finally we discuss various formation mechanisms for
  massive binaries, and the properties of massive trapezia.
  \\~\\~\\~}

\end{abstract}


\section{\textbf{INTRODUCTION}}
\label{sec:introduction}

Many reviews have been written on pre-main sequence binaries over the
past 25 years, e.g., {\em Reipurth} (1988), {\em Zinnecker} (1989),
{\em Mathieu} (1994), {\em Goodwin} (2010), and particular mention
should be made of IAU Symposium No. 200 ({\em Zinnecker and Mathieu,}
2001), which is still today a useful reference. Most recently, {\em
  Duch\^ene and Kraus} (2013) review the binarity for stars
of all masses and ages.

Stimulated by the growing discoveries of multiple systems among young
stars, there is increasing interest in the idea, first formulated by
{\em Larson} (1972), that all stars may be born in small multiple
systems, and that the mixture of single, binary, and higher-order
multiples we observe at different ages and in different environments,
may result from the dynamical evolution, driven either internally or
externally, of a primordial population of multiple systems. While more
work needs to be done to determine the multiplicity of newborn
protostars, at least -- as has been widely accepted for some time --
binarity and multiplicity is clearly established as the principal
channel of star formation. The inevitable implication is that
dynamical evolution is an essential part of early stellar evolution.
In the following we explore the processes and phenomena associated
with the early evolution of multiple systems, with a particular
emphasis on triple systems.

\section{\textbf{PHYSICS OF MULTIPLE STAR FORMATION}}
\label{sec:physics}

The collapse and fragmentation of molecular cloud cores ({\em Boss and
  Bodenheimer}, 1979) is generally agreed to be the mechanism most
likely to account for the formation of the majority of binary and
multiple star systems. 
Major advances in our physical understanding of the fragmentation
process have occurred in the last decade as a result of the
availability of adaptive mesh refinement (AMR) hydrodynamics (HD)
codes, which allow the computational effort to be concentrated where
it is needed, in regions with large gradients in the physical
variables. Many of these AMR codes, as well as smoothed particle
hydrodynamics (SPH) codes with variable smoothing lengths, have been
extended to include such effects as radiative transfer (RHD) and
magnetic fields (MHD), allowing increasingly realistic three
dimensional (3D) numerical models to be developed. We concentrate here
on the theoretical progress made on 3D models of the fragmentation
process since {\em Protostars and Planets V} appeared in 2007.

In {\em Protostars and Planets V}, the focus was on purely
hydrodynamical models of the collapse of turbulent clouds initially
containing many Jeans masses, leading to abundant fragmentation and
the formation of multiple protostar systems and protostellar clusters
({\em Bonnell et al.}, 2007; {\em Goodwin et al.}, 2007; {\em
  Whitworth et al.}, 2007). 3D HD modeling work has continued on
initially turbulent, massive clouds, with an eye toward determining
cluster properties such as the initial mass function (e.g., {\em Clark
  et al.}, 2008; {\em Offner et al.}, 2009) and the number of brown
dwarfs formed (e.g., {\em Bonnell et al.}, 2008; {\em Bate}, 2009a,b;
{\em Attwood et al.}, 2009).  3D HD SPH calculations by {\em Bate}
(2009a) made predictions of the frequency of single, binary, triple
and quadruple star systems formed during the collapse of a highly
unstable cloud with an initial mass of 500 $M_\odot$, a Jeans mass of
1 $M_\odot$, and a turbulent, high Mach number (13.7) velocity field.
This simulation involved a sufficiently large population of stars and
brown dwarfs (1250) so as to provide an excellent basis for comparison
with observed multiple systems. It is remarkable that this simulation
-- which clearly omits important physical ingredients such as magnetic
fields and radiative feedback -- nevertheless results in a reasonable
match to a wide range of observed binary parameters.  In parallel with
this study of binarity within the context of cluster formation, other
groups have instead pursued high resolution core scale simulations of
HD collapse of much lower mass, initially Bonnor-Ebert-like clouds,
delineating how factors such as the initial rotation rate,
metallicity, turbulence, and density determine whether the cloud forms
a single or multiple protostar system (see {\em Arreaga-Garc\'ia et
  al.}, 2010 and {\em Walch et al.}, 2010 for SPH and {\em Machida},
2008 for AMR calculations of this type).
 
\begin{figure}[h!]
\epsscale{0.9}
\plotone{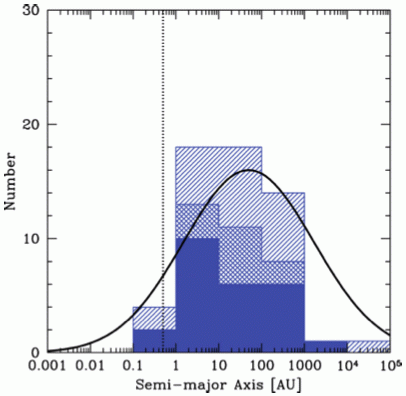}
\caption{\small Distributions of semi-major axes for primaries with
  masses greater than 0.1 $M_\odot$ (histogram) from {\em Bate} (2012), 
  compared to observations (solid line, {\em Raghavan et al.}, 2010). Solid,
  double-hatched, and single-hatched histograms are for binaries,
  triples and quadruples, respectively. The vertical line is the 
  resolution limit of the SPH calculation.
\label{fig:alan1}
}
\end{figure}

Despite this striking agreement between the outcomes of the simplest
barotropic models and observations, it is nevertheless essential to
conduct simulations that incorporate a more realistic set of physical
processes. {\em Offner et al.}  (2009) found that radiative feedback
in 3D RHD AMR calculations could indeed have an important effect on
stellar multiplicity, primarily by reducing the number of stars
formed. They also emphasized ({\em Offner et al.}, 2010) that the
inclusion of radiative feedback changes the dominant mode of
fragmentation: with a barotropic equation of state, fragmentation
normally occurs at the point when the flow is centrifugally supported
-- i.e., when it collapses into a disk at radii $< 100$~AU. This mode is
relatively suppressed when radiative feedback is included and the
fragments mainly form from turbulent fluctuations within the natal
core, at separations $\sim 1000$~AU.  Such initially wide pairs,
however, spiral in to smaller separations, an effect also found in the
simulations of {\em Bate} (2012) which are the radiative counterparts
of the previous ({\em Bate}, 2009a) calculations (see also {\em Bate},
2009b).  The resulting binary statistics are scarcely distinguishable
from those in the earlier barotropic calculations and again in good
agreement with observations (see Figure~\ref{fig:alan1}).
 
Observations of molecular clouds have shown that magnetic fields are
generally more dynamically important than turbulence, but are only one
source of cloud support against gravitational collapse for cloud
densities in the range of $10^3$ to $10^4$ cm$^{-3}$ ({\em Crutcher},
2012). While it has long been believed that magnetic field support is
lost through ambipolar diffusion, leading to gravitational collapse,
current observations do not support this picture ({\em Crutcher},
2012), but rather one where magnetic reconnection eliminates the
magnetic flux that would otherwise hinder star formation ({\em
  Lazarian et al.}, 2012). 3D MHD calculations of collapse and
fragmentation have become increasingly commonplace, though usually
assuming ideal magnetohydrodynamics (i.e., frozen-in fields) rather
than processes such as ambipolar diffusion or magnetic reconnection.

{\em Machida et al.} (2008) found that fragmentation into a wide
binary could occur provided that the initial magnetic cloud core
rotated fast enough, while close binaries resulted when the initial
magnetic energy was larger than the rotational energy. {\em Hennebelle
  and Fromang} (2008) and {\em Hennebelle and Teyssier} (2008) found
that initially uniform density and rotation magnetic clouds could
fragment if a density perturbation was large enough (50\%
amplitude), as in the standard isothermal test case of {\em Boss and
  Bodenheimer} (1979). {\em Price and Bate} (2007), {\em B\"urzle et
  al.} (2011), and {\em Boss and Keiser} (2013) all studied the
collapse of initially spherical, 1 $M_\odot$ magnetic cloud cores,
with uniform density, rotation, and magnetic fields, the MHD version
of {\em Boss and Bodenheimer} (1979). They found that clouds could
collapse to form single, binary, or multiple protostar systems,
depending on such factors as the initial magnetic field strength and
its orientation with respect to the rotation axis. When fragmentation
did occur, binary star systems were the typical outcome, along with a
few higher order systems. {\em Joos et al.} (2012) found that the
initial direction of the magnetic field with respect to the rotation
axis had an important effect on whether the collapse produced a
protostellar disk that might later fragment into a multiple system.

Radiative transfer effects were included in the models of {\em
  Commercon et al.}  (2010), who studied the collapse of 1~$M_\odot$
clouds with an AMR RMHD code, finding that frozen-in fields always
inhibited cloud fragmentation.  {\em Boss} (2009) used a 3D pseudo-MHD
code with radiative transfer in the Eddington approximation to study
the collapse and fragmentation of prolate and oblate magnetic clouds,
including the effects of ambipolar diffusion, finding that the oblate
clouds collapsed to form rings, susceptible to subsequent
fragmentation, while prolate clouds collapsed to form either single,
binary, or quadruple protostar systems. {\em Kudoh and Basu} (2008,
2011) also included ambipolar diffusion in their true MHD models,
finding that collapse could be accelerated by supersonic turbulence.

There has also been progress in adding magnetic fields and feedback
from radiation and outflows into simulations of more massive clouds,
although many such simulations do not resolve fragmentation on scales
less than $\sim 100$ AU (e.g., {\em Krumholz et al.}, 2012; {\em
  Hansen et al.}, 2012) and are thus not the simulations of choice for
following binary formation.  {\em Hennebelle et al.}  (2011) followed
the collapse of a 100 $M_\odot$ cloud with their AMR MHD code, and
found that the magnetic field could reduce the degree of
fragmentation, compared to a nonmagnetic cloud collapse, by as much as
a factor of two. {\em Commercon et al.}  (2011) extended their
previous work on 1~M$_\odot$ clouds to include 100 $M_\odot$ clouds,
but again found that magnetic fields and radiative transfer combined
to inhibit fragmentation. {\em Seifried et al.}  (2012) found that
their 100 $M_\odot$ turbulent, magnetic clouds collapsed to form just
a relatively small number of protostars.  Likewise the high resolution
simulations of {\em Myers et al.} (2013) (which combine the inclusion
of radiative transfer and magnetic fields with a resolution of $10$ AU
within a $1000~M_\odot$ cloud) find that these effects in combination
strongly suppress binary formation within the cloud.

While powerful theoretical tools now exist, along with widespread
access to large computational clusters, the huge volume of parameter
space that needs to be explored 
has to date prevented a
comprehensive theoretical picture from emerging. Nevertheless, it is
clear that in spite of the various magnetic field effects, MHD
collapse and fragmentation remains as a possibility in at least some
portions of the parameter space of initial conditions. When
fragmentation does occur in the collapse of massive, magnetic clouds,
relatively small numbers of fragments are produced, compared to the
results of 3D models of non-magnetic, often turbulent collapse, where
much larger numbers of fragments tend to form (e.g., {\em Bonnell et
  al.}, 2007; {\em Whitworth et al.}, 2007).  While such massive
clouds might form small clusters of stars, low-mass magnetized clouds
are more likely to form single or binary star systems.

In summary, then, it is premature to draw definitive conclusions about
the conditions required to produce a realistic population of binary
systems. Those simulations that can offer a statistical ensemble of
binary star systems for comparison with observations are able to match
the data very well, regardless of whether thermal feedback is employed
({\em Bate}, 2009a, 2012). The thermal feedback in these latter
simulations is, however, under-estimated somewhat ({\em Offner et
  al}., 2010) and so represents an interim case between the full
feedback and no feedback case. It remains to be seen whether
simulations with magnetic fields and full feedback do an equally good
job at matching the binary statistics, despite the indications from
the studies listed above that these effects tend to suppress binary
fragmentation.

\section{DEFINITION OF MULTIPLICITY}
\label{sec:definition}

In order to discuss observational results and compare the multiplicity
for different evolutionary stages and/or in different regions, we need
simple  and precise  terminology. Following  {\em Batten}  (1973), the
fractions  of systems  containing  exactly $n$  stars  are denoted  as
$f_n$.  The {\em multiplicity frequency} or 
{\em multiplicity fraction} $MF = 1 - f_1 = f_2 + f_3 +f_4
+  \ldots$ gives  the fraction  of non-single  systems  in a  given
sample.  This is more commonly written

\[MF = \frac{B+T+Q}{S+B+T+Q} \]   

\noindent
where S, B, T, Q are the number of single, binary, triple, and
quadruple, etc systems ({\em Reipurth and Zinnecker}, 1993).

Another common characteristic of multiplicity, the {\em companion
  star fraction} $CSF = f_2 + 2  f_3 + 3 f_4 + \ldots$ quantifies the
average number  of stellar  companions per system; it is commonly written 

\[CSF = \frac{B+2T+3Q}{S+B+T+Q} \]

\noindent
which is the average number of companions in a population, and in
principle can be larger than 1 (e.g., {\em Ghez et al.}  1997).
Measurements of $MF$ are less sensitive to the discovery of all
sub-systems than $CSF$, explaining why $MF$ is used more frequently in
comparing theory with observations. The fraction of higher-order
multiples is simply $HF = 1 - f_1 - f_2 = f_3 + f_4 + \ldots$.

\begin{figure}[ht]
\epsscale{0.59}
\plotone{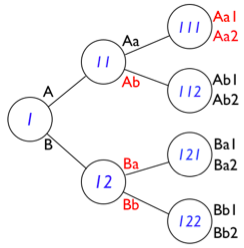}
\caption{\small The structure  of  hierarchical   multiple  systems
    can be represented  by a
    binary graph, the figure describes all possible multiples 
    up to an octuple system.   The position of  each sub-system is
    coded by {\em levels} shown in blue in the circles. An example of a 
    pentuple system is marked with red letters, the outer pair A,B is
    at level {\em 1}, the innermost sub-sub-system Aa1,Aa2 is 
    at level {\em 111}. The nomenclature follows the IAU recommendation.
\label{fig:levels}
}
\end{figure}

The vast majority of observed multiple systems are {\em hierarchical}:
the ratio of separations between their inner and outer pairs is large,
ensuring long-term dynamical stability.  Stellar motions in stable
hierarchical systems are represented approximately by Keplerian
orbits.  Hierarchies can be described by binary graphs or trees
(Figure~\ref{fig:levels}). The position of each sub-system in this graph
can be coded by its {\em level}.  The outermost (widest) pair is at
the root of the tree (level {\em 1}). Inner pairs associated with
primary and secondary components of the outer pair are called levels
{\em 11} and {\em 12}, respectively, and this notation continues to
deeper levels. Triple systems can have inner pairs at level {\em 11} or
{\em 12}.  When both sub-systems are present, we get the so-called
{\em 2+2 quadruple}.  Alternatively, a {\em planetary} quadruple
system consists of levels {\em 1}, {\em 11}, and {\em 111}; it has two
companions associated with the same primary star.

In principle, the most precise description of multiplicity statistics
would be the joint distribution of the main orbital parameters (period
or semi-major axis, mass ratio, and eccentricity) at all hierarchical
levels. But even for simple binaries such a 3-dimensional distribution is
poorly known, and the number of variables and complexity increases quickly
when dealing with triples, quadruples, etc.  To first order, the
multiplicity is characterized by the fractions $f_n$ or by their
combinations such as $MF$ (which equals the fraction of level-{\em 1}
systems), $CSF$, and $HF$.

\begin{figure*}[tbp]
\epsscale{1.0}
\plotone{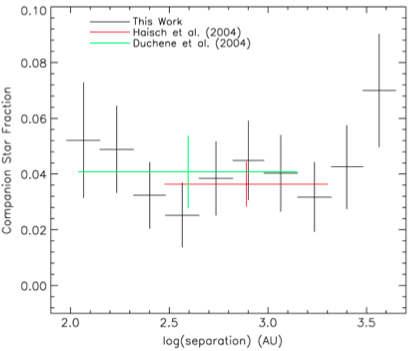}
\epsscale{0.85}
\plotone{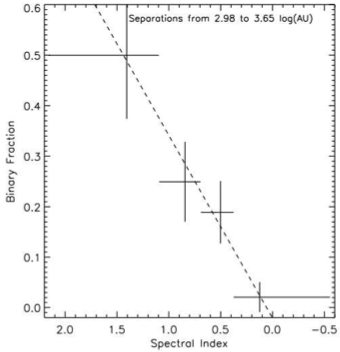}
\caption{\small
(a): The separation distribution function of embedded
  protostellar binaries. There is a strong excess of widely separated
  companions with separations larger than 1000~AU. ~~(b): The
  population of wide companions is found to disappear with decreasing
  spectral index, which is a proxy for age.  From {\em Connelley et al.} (2008a,b).
\label{fig:connelley}
}
\end{figure*}

\section{\textbf{OBSERVATIONS OF PROTOSTELLAR BINARIES AND MULTIPLES}} 
\label{sec:proto}

Studies of binaries and multiples during the protostellar stage are
important, since they offer the best chance of seeing the results of
fragmentation of molecular clouds, as discussed in
Section~\ref{sec:physics}.  However, most protostars are still
deeply embedded, so such observations are hampered by extinctions that
can exceed A$_V$$\sim$ 100~mag. Hence, infrared, submillimeter, or
radio continuum observations are required.

\subsection{Infrared Observations }
\label{sec:protoIR}

Class~I protostars are often detectable at near-infrared wavelengths,
although for disk orientations near edge-on one sees them only in
scattered light. In contrast, the massive circumstellar environment of
Class~0 sources make them detectable only at longer wavelengths.  {\em
  Haisch et al.} (2004) performed a near-infrared imaging survey of 76
Class~I sources and found a companion star fraction of 18\%$\pm$4\% in the
separation range $\sim$300-2000~AU. 
In a similar study, {\em Duch\^ene et al.} (2004) obtained a companion
star fraction of 27\%$\pm$6\% in the range 110-1400~AU. To detect
closer companions, {\em Duch\^ene et al.} (2007) used adaptive optics
to survey 45 protostars, and found a companion star fraction of
47\%$\pm$8\% in the range 14-1400~AU; comparison of the two numbers
indicate the prevalence of close protostellar companions. In a major
survey of 189 Class~I sources, {\em Connelley et al.} (2008a,b)
detected 89 companions, and the separation distribution function is
shown in Figure~\ref{fig:connelley}a.  For the closer separations, it
is seen to be very similar to that of T~Tauri binaries. But for larger
separations, a clear excess of wide companions (with separations up to
4500~AU) becomes evident, which is not seen for the more evolved
T~Tauri stars. When plotting the binary fraction as a function of
spectral index, which measures the amount of circumstellar material
and is used as a proxy for stellar age, they find a dramatic decline
in these wide companions (Figure~\ref{fig:connelley}b), from
$\sim$50\% to $<$5\%. In other words, powerful dynamical processes
must occur during the Class~I phase, leading to the dispersal of a
significant population of wide companions.  These observations can be
understood in the context of the dynamical evolution of newborn
multiple systems, which in most cases break up, leading to the
ejection of one of the components (see Section~\ref{sec:dynamics}).
Such ejected components should be observable for a while, and {\em
  Connelley et al.} (2009) found that of 47 protostars observed with
adaptive optics, every target with a close companion has another young
star within a projected separation of 25,000~AU.

The study of even closer companions to protostars is still in its infancy.
In a pilot program, {\em Viana Almeida et al.} (2012) found large radial
velocity variations in three out of seven embedded sources in
Ophiuchus, and speculated that they could be evidence for
spectroscopic protobinaries. {\em Muzerolle et al.} (2013) used the
Spitzer Space Telescope to monitor IC~348 and found a protostar
showing major luminosity changes on a period of 25.34~days; the most
likely explanation is that a companion in an eccentric orbit drives
pulsed accretion around periastron.

\subsection{Submillimeter Observations}\
\label{sec:protoSUBMM}

\vspace{-0.5cm}

\begin{figure}[t]
\epsscale{1.00}
\plotone{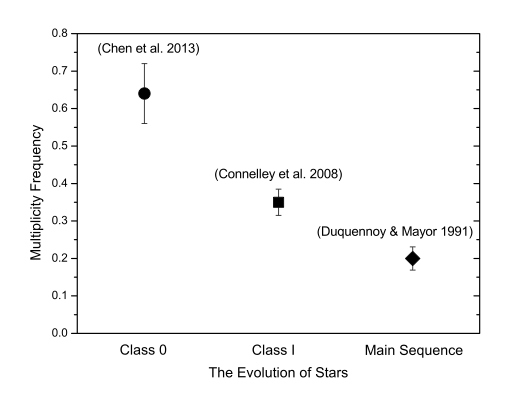}
\vspace{-0.9cm}
\caption{\small The multiplicity frequency declines through the
  protostellar phase because of the breakup of small multiple systems.
From {\em Chen et al.} (2013).
\label{fig:chen}
}
\end{figure}

The study of binarity of the youngest protostars, the Class~0 sources,
requires longer wavelength observations, and mm interferometry has
become a powerful tool to study binarity of protostars. Pioneering
work was done by {\em Looney et al.} (2000), who observed 7 Class~0
and I sources; further small samples of embedded sources were observed
by, e.g., {\em Chen et al.} (2008, 2009), {\em Maury et al.} (2010),
{\em Enoch et al.} (2011), and {\em Tobin et al.} (2013). All of these
studies suffer from very small samples, and hence yield uncertain
statistics. This problem has been alleviated by {\em Chen et al.}
(2013), who presented high angular resolution 1.3~mm and 850~$\mu$m
dust continuum data from the Submillimeter Array for 33 Class~0
sources. No less than twenty-one of the sources show evidence for
companions in the projected separation range from 50 to 5000~AU.  This
leads to a multiplicity frequency MF = 0.64$\pm$0.08 and a companion
star fraction CSF = 0.91$\pm$0.05 for Class~0 protostars. As noted by
{\em Chen et al.} (2013), their survey is complete for systems larger
than $\sim$1800~AU, and hence these values must be regarded as lower
limits. Given that numerous Class~0 binaries may have much closer
companions, these results are consistent with the possibility that
virtually all stars are born as binaries or multiples, an idea that
dates back to {\em Larson} (1972).  Figure~\ref{fig:chen} shows in
graphical form the observed decrease in binarity as a function of
evolutionary stage, a result that strongly supports a view of early
stellar evolution in which small multiple systems evolve dynamically,
break up, and the decay products eventually evolve into the
distribution of singles, binaries, and higher-order multiples we
observe in the field.

\subsection{Radio Continuum Observations}
\label{sec:protoVLA}

As is clear from the discussion above, it is critically important to
study protostars with much higher resolution in order to determine the
multiplicity at small separations.  Radio observations are the only
technique available at present that allows the study with high angular
resolution of the earliest stages of star formation. These studies can
be performed with an angular resolution of order 0.1 arcsec with radio
interferometers such as the Jansky Very Large Array (VLA) and the
expanded Multi-Element Radio-Linked Interferometer Network (eMERLIN).
What is detected here is the free-free emission from the base of the
ionized outflows that are frequently present at early
evolutionary stages. These structures trace the star with high
precision, and favors the detection of very young 
Class 0, I, and II objects. 

A series of VLA studies (e.g., {\em Rodr\'\i guez et al.}, 2003, 2010;
{\em Reipurth et al.}, 2002, 2004) show binary and multiple sources
clustered on scales of a few hundred AU. 
A binary frequency of order $\sim$33\% is found in these studies.
Since not all sources show free-free emission, and those which do are
often found to be variable, such statistics provide only lower limits.

If the star has strong magnetospheric activity, the resulting
gyrosynchrotron emission is compact and intense enough to be observed
with the technique of Very Long Baseline Interferometry (VLBI) that
can reach angular resolutions of order 1 milliarcsecond and better,
and that allows the study of stellar motions with great detail. This
technique favors the detection of the more evolved class III stars.
It should be noted, however, that at least one class I protostar, IRS
5b in Corona Australis, has been detected with VLBI techniques ({\em Deller
et al.}, 2013).  In a series of studies to determine the parallax of
young stars in Gould's Belt ({\em Loinard}, 2013), it has been found that
several are binary and it has been possible to follow their orbital
motions (e.g., {\em Torres et al.}, 2012) and to study the
radio emission as a function of separation, finding evidence of
interaction between the individual magnetospheres.  Radio emission of
non-thermal origin has been detected all the way down to the ultracool
dwarfs (late M, L, and T types), in some sources in the form of
periodic bursts of extremely bright, 100\% circularly polarized,
coherent radio emission (e.g., {\em Hallinan et al.}, 2007). 

With the new generation of centimeter and millimeter interferometers,
especially ALMA, the field of radio emission from binary and multiple
young stellar systems faces a new era of opportunity that should
result in much better statistics, especially in the protostellar
stage.

\section{\textbf{DYNAMICS OF MULTIPLE STARS}}
\label{sec:dynamics}

\begin{figure}[t]
\epsscale{0.97}
\plotone{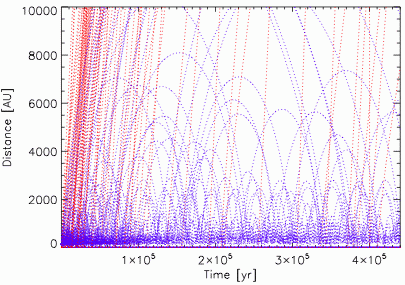}
\caption{\small 100 simulations showing the dynamical evolution of a triple
  system of three 0.5~M$_\odot$ stars with initial mean separations of
  100~AU embedded in a 3~M$_\odot$ cloud core. Many of the initial ejections are escapes, but the majority fall back, to be ejected sometimes again and again.
From {\em Reipurth et al.} (2010).
\label{fig:100sim}
}  
\end{figure}

If three bodies are randomly placed within a volume, then more than
98\% of the systems will be in a non-hierarchical configuration, that
is, the third body is closer than $\sim$10 times the separation of the
other two bodies.  It is well known that such configurations are
inherently unstable, and will on a timescale of around 100 crossing
times decay into a hierarchical configuration, in a process where the
third body is ejected, either into a distant orbit or into an escape,
see Figure~\ref{fig:100sim}, (e.g., {\em Anosova}, 1986; {\em Sterzik and
Durisen}, 1998; {\em Umbreit et al.}, 2005).  The energy to do this
comes from the binding energy of the remaining binary, which as a
result shrinks and at the same time frequently gets a highly eccentric
orbit. For such an ejection to take place, the three bodies must first
meet in a close triple approach, during which energy and momentum can
be exchanged.  A detailed analysis of the dynamics of triple systems
can be found in {\em Valtonen and Karttunen} (2006).

\begin{figure}[t]
\epsscale{0.9}
\plotone{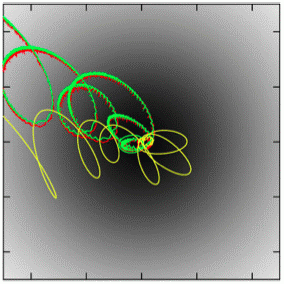}
\caption{\small An example of the chaotic orbits of three bodies born
  in and accreting from a cloud core. The triple system in this simulation 
  remains bound as it drifts away from the core, but is unstable and will
  eventually break apart. The figure is 10,000~AU across.  
  From {\em Reipurth and Mikkola} (2014).
\label{fig:orbitcore}
}
\end{figure}

N-body simulations that include the potential of a cloud core reveal
that many systems break up shortly after formation, sending the third
body into an escape, but the majority goes through several or many
ejections that are too weak to escape the potential well, and the
third body thus falls back ({\em Reipurth et al.}, 2010).  As the
cloud core gradually shrinks through accretion, outflows, and
irradiation, the third body eventually manages to escape.  In some
cases the triple remains bound until after the core has disappeared
(see Figure~\ref{fig:orbitcore}), but only about $\sim$10\% of triples
are stable enough to survive on long timescales. The body that is
ejected is most often the lowest-mass member, but complex dynamics can
lead to many other configurations and outcomes. Stochastic events play
an essential but unpredictable role in the early stages of triple
systems, and so their evolution can only be understood statistically.

\begin{figure}[t]
\epsscale{1.0}
\plotone{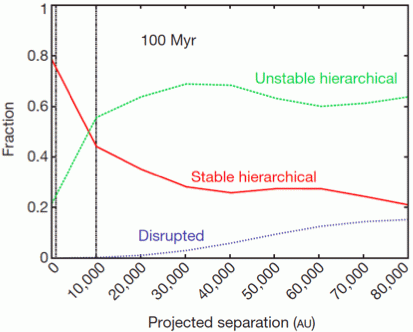}
\caption{\small The relative numbers of bound stable, bound unstable,
  and unbound triple systems as function of the projected
  separation of the outer pairs, from a major N-body simulation 
  after 100~Myr. The
  majority of very wide binaries is unstable at this age. At much
  younger ages, say 1~Myr, unstable systems dominate at all
  separations.  (From {\em Reipurth and Mikkola}, 2012).
  \label{fig:relativenumbers} 
}
\end{figure}

A stability analysis of hierarchical bound triple systems formed in
N-body simulations shows that they divide into stable and unstable
systems. Any time that a distant third component passes through a
periastron passage and comes close to the inner binary, there is the
possibility of an instability of the system, depending on the
configuration of the inner binary.  Stable triples remain bound for
hundreds of millions or billions of years, but unstable systems can
break apart at any time.  Figure~\ref{fig:relativenumbers} shows the
fraction of triples that after 100~Myr are stable, unstable, or
already disrupted, as function of their projected separation, from a
major N-body simulation. For separations less than 10,000~AU (vertical
line) the majority is stable, but for wider separations unstable
systems dominate. For young systems in star-forming regions, however,
unstable systems significantly dominate at all separations. These
unstable systems will soon break apart. For young ages, one therefore
observes many more triple systems than at older ages ({\em Reipurth
  and Mikkola}, 2012).

\begin{figure}[th!]
\epsscale{0.90}
\plotone{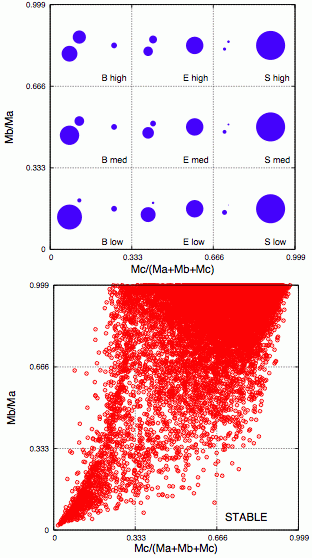}
\caption{\small {\em (a)}: Location and definition of nine different types 
  of triple systems in the triple diagnostic diagram.  
  ~~~~{\em (b)}: The location of 15,524
  stable triple systems in the triple diagnostic diagram at an age of
  1~Myr. Since all triple systems in these simulations were started
  out with three identical bodies, the original systems were all
  located at the point (0.333,1.000). Their final location is
  determined by their dynamical evolution and resulting accretion.
  From {\em Reipurth and Mikkola} (2014).  
\label{fig:triplediagram}
}
 \end{figure}

Triple systems can be classified in the {\em triple diagnostic
  diagram} (Figure~\ref{fig:triplediagram}a), where the mass ratio of
the binary is plotted as a function of the mass of the third body
relative to the total system mass ({\em Reipurth and Mikkola}, 2014).
In the right hand of the diagram reside the systems that are dominated
by a massive single star ({\em S-type}), to the left are those where a
massive binary dominates ({\em B-type}) and in the middle are the
systems where the mass is about equally distributed in the binary and
the single ({\em E-type}).  Sub-divisions can additionally be made
depending on the mass ratio of the binary (high, medium, low).  Note
that since the axes represent ratios, i.e., dimensionless numbers, then
the absolute mass of the system is not involved.  This simple
classification system encompasses all categories of triple systems. As
the name indicates, the distribution of systems in the diagram harbors
important diagnostics for understanding the early evolution of triple
systems.  Figure~\ref{fig:triplediagram}b shows the result of N-body
calculations that include accretion as the three bodies move around
each other inside the cloud core. All systems in the diagram are
long-term stable. To better isolate the interplay between dynamics and
accretion, all three components started out with equal masses, i.e.
they were initially placed at (0.333, 1.000). As is evident, the
interplay between dynamics and accretion can lead to very different
outcomes, with some areas of the diagram populated much more densely
than others. Comparison with complete, unbiased samples of triples
will provide much insight into the formation processes of triple
systems.

\subsection{Origin of Brown Dwarf/VLM Binaries}
\label{sec:pmsVLM}

The formation of very low mass (M$\lesssim$0.1~M$_\odot$) objects has
been debated for a long time, and three basic ideas have emerged: a
very low mass (VLM) object can form if the nascent core has too little
mass ({\em Padoan and Nordlund}, 2004), or it can form if the stellar
seed is removed from the infall zone through dynamical ejection ({\em
  Reipurth and Clarke}, 2001; {\em Stamatellos et al.}, 2007; {\em
  Basu and Vorobyov}, 2012), or the cloud core can photoevaporate if a
nearby OB star is formed ({\em Whitworth and Zinnecker}, 2004). The
emerging consensus is that all three mechanisms are likely to operate
under different circumstances, and that the relevant question is not
which mechanism is correct, but how big their relative contributions
are to the production of VLM objects ({\em Whitworth et al.}, 2007).
Similarly, BD/VLM binaries are likely to have several formation
mechanisms.

Extensive numerical studies combining N-body simulations with
accretion have shown that the large majority of brown dwarf ejections
are not violent events, but rather the result of unstable triple
systems that eventually drift apart at very small velocities,
typically within the first 100~Myr ({\em Reipurth and Mikkola}, 2014).
When brown dwarfs are released from triple systems, by far the
majority of the remaining binaries are VLM objects. These binaries
gently recoil and become isolated VLM binaries. These VLM binaries
have a semimajor axis distribution that peaks at around 10-15~AU, but
with a tail stretching out to $\sim$250~AU. At shorter separations,
the simulations show a steep decline in number of systems, although
the simulations underestimate the number of close binaries because
they do not take viscous orbital evolution into account. Brown dwarf
and VLM binaries formed through dynamical interactions can in
principle have much larger separations, of many hundreds or thousands
of AU, but in that case they must be bound triple systems, where one
component is a close, often unresolved, binary.  More than 90\% of
bound triple systems at 1~Myr have dispersed by 100~Myr, and {\em all}
VLM triple systems with {\em outer} semimajor axes less than a few
hundred AU have broken up. In this context, it is interesting that
{\em Biller et al.}  (2011) found an excess of 10-50~AU young brown
dwarf binaries in the 5~Myr old Upper Scorpius association compared to
the field.

\subsection{Origin of Spectroscopic Binaries}
\label{sec:pmsSPECBIN}

Spectroscopic binaries is a generic term for all binaries that have
separations so close that their orbital motion is measurable with
radial velocity techniques. In practical terms, the large majority of
known spectroscopic binaries has a period less than $\sim$4,000~days.
Surveys of metal-poor field stars (for which statistics is
particularly good) find that 18\%$\pm$4\% are spectroscopic binaries
({\em Carney et al.}, 2003). 

Radial velocity studies of young stars are complicated by the
sometimes wide and/or complex line profiles, but an increasing number
of pre-main sequence spectroscopic binaries are now known (e.g., {\em
  Mathieu}, 1994; {\em Melo et al.}, 2001; {\em Prato}, 2007; {\em
  Joergens}, 2008; {\em Nguyen et al.}, 2012).  In the Orion Nebula
Cluster, {\em Tobin et al.}  (2009) has up to now found that 11.5\% of
the observed members are spectroscopic binaries, but the survey is
still ongoing.

Spectroscopic binaries have semi-major axes that are often
measured in units of stellar radii, and rarely exceed a few AU. These
binaries cannot form with such close separations, and they must
therefore result from processes that cause a spiral-in of an initially
wider binary system.  Given that newborn binaries and multiples are
surrounded by a viscous medium during the protostellar phase,
components can naturally spiral in during the star-forming process
because of dynamical friction with the surrounding medium (e.g., {\em
  Gorti and Bhatt}, 1996; {\em Stahler}, 2010; {\em Korntreff et al.},
2012), see Section~\ref{sec:gas}.

The evolution of embedded triple systems can enhance or initiate this
process. When a newborn triple system is transformed from a
non-hierarchical configuration to a hierarchical one, the newly bound
binary shrinks in order for the third body to be ejected into a more
distant orbit or into an escape. The binary also gets a highly
eccentric orbit as a result of this process. Given that the components
are surrounded by abundant circumstellar material at early
evolutionary stages, the shrinkage and the eccentric orbits force
regular dissipative interactions, leading to orbital decay.  As
discussed by {\em Stahler} (2010), the ultimate result can be a
merger. But if the infall of new material from an envelope ceases
before that, then the orbital decay is halted, and the binary ends up
with the orbital parameters it happens to have when the viscous
interactions cease (except for very close binaries, where orbital
circularization will occur -- {\em Zahn and Bouchet}, 1989).

Spectroscopic binaries originating in a triple system therefore have
an important stochastic element in their evolution, depending on when
the triple system broke up and when circumstellar material became
exhausted.

\subsection{Origin of Single Stars}
\label{sec:pmsSINGLE}

The strong increase in number of stars (single and non-single) for
decreasing mass combined with the strong decrease in number of
binaries also for decreasing mass (see Section~\ref{sec:statistics})
led {\em Lada} (2006) to conclude
that the majority of stars in our Galaxy are single. And for
solar-type stars in the solar neighborhood, {\em Raghavan et al.}
(2010) found that 56\% are single.  This high preponderance of single
stars is not consistent with the very high multiplicity frequency
determined among protostars (see Section~\ref{sec:protoSUBMM}), and
leads to interesting questions about the origin of single stars.

When we observe a single star, it may have one of three origins:
it can be born in isolation; it may have been born in a multiple
system that decayed and ejected one component; or it may even be the
product of two stars in a binary that spiraled in and merged during
the protostellar phase. 

Historically, mergers have been considered almost exclusively in the
contexts of late stellar evolutionary stages or massive stars.
Intriguingly, new N-body simulations of low-mass small-N multiple
systems and studies of orbital decay in a viscous medium indicate that
mergers may occur in a non-negligible number of cases during early
stellar evolution ({\em Rawiraswattana et al.}, 2012; {\em Stahler},
2010; {\em Leigh and Geller}, 2012; {\em Korntreff et al.}, 2012).

Small triple systems, whether formed in isolation or in a cluster,
will evolve dynamically, and $\sim$90\% break up, each producing a
single star, which drifts away with a velocity around 1 km/sec. This
corresponds to $\sim$100,000 AU in half a million years, so very soon
such ejecta will disperse and any trace of their origin will be lost.
Because of dynamical processing, it is the lowest mass components that
tend to escape. Newborn higher-order multiples such as quadruples, pentuples,
sextuples, etc., may produce more than one single star per star
forming event. 

The formation of a single star from a collapse event is -- not
surprisingly -- the standard view of the origin of single stars.
However, the very high multiplicity of protostars (see
Section~\ref{sec:protoSUBMM}) has by now made it clear that 
single-star collapse is not
the principal channel of star formation. And it should by no means be
automatically assumed that young single stars found in a low-mass
star-forming region represent cases of single, isolated star formation.

\section{\textbf{OBSERVATIONAL CONSEQUENCES}}
\label{sec:consequences}

\begin{figure}[t!]
 \epsscale{0.62}
\plotone{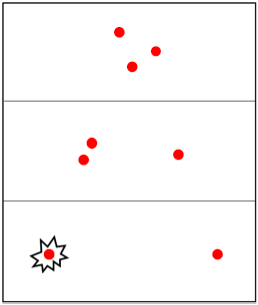}
\caption{\small A schematic plot of the evolution of a triple
  system, from non-hierarchical to hierarchical (top two panels)
  followed by the binary viscous in-spiral phase leading to disk-disk
  interactions and in some cases stellar mergers. 
\label{fig:diskcollision}
}
 \end{figure}

The dynamical evolution discussed above has observational consequences:

{\bf FUor Eruptions.}  The close triple encounters in triple systems,
that are prerequisites for the ejection of one of the components, are
statistically most likely to occur during the protostellar stage ({\em
  Reipurth}, 2000). At this stage the three bodies are surrounded by
significant amounts of circumstellar material, which will interact and
cause a major brightening, from accretion and shock-heating. These
events we here call {\em Encounters of Type~1}. After the hierarchical
configuration has been achieved, the shrinking of the binary orbit and
its high eccentricity will lead to a series of disk-disk interactions
at each periastron passage (Figure~\ref{fig:diskcollision}). The disks
will be seriously disturbed, causing eruptions, but much of the mass
will fall back and reassemble in the disk again ({\em Clarke and
  Pringle}, 1993; {\em Hall et al.}, 1996; {\em Umbreit et al.}, 2011;
see Section~\ref{sec:retention}).  As a result of this viscous
evolution, the binary shrinks until the point when the stars are so
close that the circumstellar material shifts from being in two
circum{\em stellar} disks to instead assemble in one circum{\em
  binary} disk ({\em Reipurth and Aspin}, 2004).  This sequence of
eruptions is called {\em Encounters of Type~2}.  Finally, if the
triple evolution occurs so early that abundant gas is present, then
the inspiral phase of the binary can result in the coalescence of the
two stars (e.g., {\em Stahler}, 2010; {\em Rawiraswattana et al.},
2012; {\em Leigh and Geller}, 2012); such events are called {\em
  Encounters of Type~3}.  Observations have revealed various types of
outbursts among young stars, the main one being the FUor eruptions
({\em Herbig}, 1977), see the {\em Audard et al.} chapter.  Once enough
detailed observations have become available, it may be possible to
identify those that result from triple evolution, since each of the
above types of encounters are likely to have characteristic energy
releases and timescales, which may make them identifiable. It will be
challenging to disentangle the various types of eruptions observed,
since disks obviously can be disturbed also internally through
instabilities, and disks have limited ways to react to perturbations,
whether internal or external.

{\bf Herbig-Haro Flows.} Accretion and outflow is generally coupled,
and so the abovementioned encounters will give rise to different
outflow characteristics, at young ages manifested as Herbig-Haro flows
({\em Reipurth and Bally}, 2001). Encounters of Type~1 from close
triple approaches will result in one or a few giant bow shocks, while
a sequence of Type~2 encounters will produce closely spaced knots,
driven by cyclic accretion modulated on an orbital timescale, as seen
in the finely collimated Herbig-Haro jets. Once the binary components
have spiraled in so close that disk truncation rips up the magnetic
field anchoring that supports the jet launch platform, then the
collimated outflow phase is terminated, and subsequent mass loss will
appear as massive but uncollimated winds, like those seen in the
spectra of FUor eruptions. Seen in this perspective, giant HH flows
represent a fossil record of the accretion history primarily dictated
by the orbital evolution of their driving sources, which are expected
to be multiple, as frequently observed ({\em Reipurth}, 2000).
Other disk instabilities can also form Herbig-Haro flows, but on a
smaller scale.

{\bf Orphaned Protostars.} The many dynamical ejections in which the
third body fails to escape the potential well of the core plus
remaining binary instead lead to large excursions, where the third
body for long periods is tenuously bound in the outskirts or outside
the cloud core.  If such ejections occur during the protostellar
stage, as many do, then these {\em orphaned protostars} open the
possibility to study naked protostars still high up on their Hayashi
tracks at near-infrared and even at optical wavelengths ({\em Reipurth
  et al.}, 2010). The triple system T~Tauri may be a case.

{\bf Formation of Wide Binaries.} Binaries with semimajor axes as
large as 10,000~AU are now frequently found thanks to increasing
astrometric precision. They challenge our understanding of star
formation because their separations exceed the typical size of a
collapsing cloud core. The dynamical evolution of multiple systems
offers a simple way to form such wide binaries: although born compact,
a triple system can dynamically scatter a component to very large
distances, thus unfolding the triple system into an extreme
hierarchical architecture. Many very wide binaries are therefore
likely to be triples or higher-order multiples, although true binaries
can also form when a merger has taken place in an encounter of Type~3
({\em Reipurth and Mikkola}, 2012).  Another independent mechanism
that forms wide binaries in clusters is discussed in
Section~\ref{sec:clusterENCOUNTER}.

\section{\textbf{PRE-MAIN SEQUENCE BINARIES/MULTIPLES}} 
\label{sec:pms}

It has been known since the early studies of T~Tauri stars that some
are binaries ({\em Joy and van Biesbroeck}, 1944; {\em Herbig}, 1962).
Further interest was spurred by the discovery of an infrared companion
to T~Tauri by {\em Dyck et al.} (1982). In 1993, three major surveys
appeared which established that T~Tauri stars have about twice the
binary frequency compared to field stars, at least for the wider pairs
({\em Reipurth and Zinnecker}, 1993; {\em Leinert et al.}, 1993; {\em
  Ghez et al.}, 1993). In the following, we examine the status 20
years later.

\subsection{Statistics and Environment} 
\label{sec:pmsSTATISTICS}

When comparing the multiplicity of young stars to the field, the key
reference for solar-type field stars has been {\em Duquennoy and
  Mayor} (1991). Since that study, observational techniques have
improved, and {\em Raghavan et al.} (2010) have studied 454 F6-K3
dwarf and subdwarf stars within 25~pc using many different techniques.
Their observed fractions of single, binary, triple and quadruple stars
are 56$\pm$2\%:33$\pm$2\%:8$\pm$2\%:3$\pm$1\%, yielding a
completion-corrected multiplicity frequency of 46\%, and implying that
among {\em solar type} stars, the majority are single. They also found
that 25\% of non-single stars are higher-order multiples, and that the
percentage of triple and quadruple systems is roughly twice that
estimated by {\em Duquennoy and Mayor} (1991).  Systems with larger
cross sections, i.e., those with more than two components or with long
orbital periods, tend to be younger, indicating the loss of components
with time.

{\em De Rosa et al.} (2014) have studied 435 A-type stars and, within
the errors, they find the precise same fractions of singles,
binaries, etc., as {\em Raghavan et al.} (2010) did for later-type
stars.

Among more massive stars, the radial velocity study of {\em Chini et
  al.} (2012) examined 250 O-stars and 540 B-stars and found that more
than 82\% of stars with masses above 16~M$_\odot$ form {\em close}
binaries, but that this high frequency drops monotonically to less
than 20\% for stars of 3~M$_\odot$ (see Section~\ref{sec:massive}).
For late type stars, {\em Fischer and Marcy} (1992) found a binary
frequency of 42$\pm$9\% among nearby M dwarfs, while {\em Bergfors et
  al.} (2010) for M0-M6 dwarfs measured 32$\pm$6\% in the range
3-180~AU.  For very late-type stars (M6 and later) {\em Allen} (2007)
determined a binary frequency of 20-22\%, consistent with the
$\sim$24\% binary frequency found for L dwarfs by {\em Reid et al.}
(2006).  And {\em Kraus and Hillenbrand} (2012) found a smooth decline
in binary frequency from 0.5~M$_\odot$ to 0.02~M$_\odot$.  Altogether,
these results confirm the trends seen in various other investigations
(e.g., {\em Raghavan et al.}, 2010), namely that {\em binarity is a
  strongly decreasing function with decreasing stellar mass}.

For young stars, getting good statistics is obviously more difficult.
The more massive young stars, the Herbig Ae/Be stars, have long been
known to have a high binarity. {\em Leinert et al.} (1997) used
speckle interferometry to find a binary frequency of 31\% to
42\%, while {\em Baines et al.} (2006) used spectro-astrometry to
determine a binary frequency of 68$\pm$11\%, with a hint that the
binarity of Herbig Be stars is higher than for the Herbig Ae stars.
To this should be added the spectroscopic binaries, which {\em
  Corporon and Lagrange} (1999) found to be around 10\%. {\em
  Kouwenhoven et al.} (2007) analyzed several data sets on the Upper
Sco association, and found that intermediate mass stars have a
binary frequency $>$70\% at a 3$\sigma$ confidence level. 

The most thoroughly examined low-mass star-forming region is
Taurus-Auriga, and in a detailed study  {\em Kraus et
  al.} (2011) found an observed multiplicity frequency of $\sim$60\%
for separations in the range 3--5000~AU. When corrections are
done to account for missing very close and very wide companions,
the multiplicity frequency rises to $\sim$67--75\%.

Taurus-Auriga, however, appears to be different from other low mass
star-forming regions (e.g., {\em Correia et al.}, 2006), see
Section~\ref{sec:clusterOBS}. Chamaeleon~I was studied by {\em
  Lafreni\`ere et al.} (2008), who found a multiplicity frequency of
30$\pm$6\% over the interval $\sim$16-1000~AU.  In Ophiuchus, {\em
  Ratzka et al.}  (2005) determined a multiplicity frequency of
29$\pm$4\% in the range 18--900~AU, while in the Upper Scorpius region
of the Scorpius-Centaurus OB association {\em Kraus et al.} (2008)
found a binary frequency of 35$\pm$5\% in the 6-435~AU range. When
properly scaled and compared, these values are consistent within their
errors, suggesting that Taurus is atypical.

Other observations indicate that multiplicity differs among some
regions. For example, {\em Reipurth and Zinnecker} (1993) found that
young stars in clouds with ten or fewer stars were twice as likely to
have a visual companion as clouds with more stars.  {\em Brandeker et al.}
(2006) found a deficit of wide binaries in the $\eta$ Chamaeleontis
cluster.  {\em Kraus and Hillenbrand} (2007) noted that the wide
binary frequencies in four star-forming regions are dependent on both
the mass of the primary star and the environment, but did not find a
relation with stellar density.  {\em Connelley et al.} (2008b)
found that the binary separation distribution of Class~I sources in a
distributed population in Orion (not near the Orion Nebula Cluster) is
significantly different from other nearby, low-mass star-forming
regions.

Naturally, these results raise the question whether the environment
plays a role for the population of binaries and multiples. It is
conceivable that different physical conditions can affect the
frequency and properties of newborn binaries ({\em Durisen and
  Sterzik}, 1994; {\em Sterzik et al.}, 2003). And longer-term
dynamical interactions between binaries and single stars will depend
on the stellar density in the birth environment (e.g., {\em Kroupa},
1998; {\em Kroupa and Bouvier}, 2003; {\em Kroupa and Petr-Gotzens},
2011).  Assuming all stars are formed as binaries in groups and
clusters of different densities, {\em Marks and Kroupa} (2012) show
that -- using an inverse dynamical population synthesis -- the
abovementioned binary properties in different star-forming regions can
be reproduced.  This is further discussed in
Section~\ref{sec:cluster}.

\subsection{The Separation Distribution Function}
\label{sec:pmsSEPDIST}

Binaries have separations spanning an enormous range, from contact
binaries to tenuously bound ultrawide binaries and proper motion pairs
with separations up to a parsec (and possibly even more). The way binaries
are distributed along this vast range in separations carries information
on both the mechanisms of formation and subsequent dynamical (and
sometimes viscous) evolution. We note that almost all authors for
practical reasons use projected separations. Because most binaries are
eccentric and therefore spend more time near apastron than at
periastron, one can show that -- for reasonable assumptions about
eccentricity -- the mean instantaneous projected separations and mean
semimajor axes differ by only $\sim$5\% (e.g., {\em van Albada}, 1968).

{\em \"Opik} (1924) suggested that the distribution of separations for
field binaries follows a log-flat distribution {\em f(a) $\propto$
  1/a}, whereas {\em Kuiper} (1942) found a log-normal distribution;
the latter has been supported by both {\em Duquennoy and Mayor} (1991)
and {\em Raghavan et al.} (2010), who found the peak of the
distribution of solar-type binaries to be at $\sim$30~AU and
$\sim$50~AU, respectively, and \"Opik's Law is no longer considered
for closer binaries. But the distribution of the widest binaries can
be fitted with a power law, although with an exponent between $-$1.5
and $-$1.6, decreasing somewhat faster than \"Opik's Law ({\em
  L\'epine and Bongiorno}, 2007; {\em Tokovinin and L\'epine}, 2012).

For young low-mass stars the separation distribution function is less
well known. For clusters, the absence of wide binaries has been noted
in the Orion Nebula Cluster ({\em Scally et al.}, 1999; {\em Reipurth
  et al.}, 2007), see Section~\ref{sec:clusterOBS}. Among less densely
populated low-mass star-forming regions, the most detailed study is of
Taurus by {\em Kraus et al.} (2011). They find that the separation
distribution function for stars in the mass range from 0.7 to
2.5~M$_\odot$ is nearly log-flat over the wide separation range 3--5000~AU,
that is, there are relatively more wide binaries among young stars
than in the field.  

For very low mass (VLM) objects, it has been known for some time that the mean
separation and separation range of binaries, both young and old,
shrink with decreasing mass (see {\em Burgasser et al.}, 2007 and
references therein).  Where {\em Fischer and Marcy} (1992) found that
M-star binary separations peak around 4-30~AU, {\em Burgasser et al.}
(2007) estimated that VLM objects peak around $\sim$3-10~AU.  {\em
  Kraus and Hillenbrand} (2012) studied low-mass (0.02--0.5~M$_\odot$)
young stars and brown dwarfs in nearby associations and found that the
mean separation and separation range of binaries decline
smoothly with mass; a degeneracy between total binary frequency and
mean binary separation, however, precludes a more precise description
of this decline.

\subsection{Mass Ratios}
\label{sec:pmsMR}

The mass ratios ($q = M_2/M_1$) we observe for young stars are
dominated by processes during the protobinary accretion phase, and
subsequent circumbinary disk accretion will have only limited effect
on the mass ratios (see Section~\ref{sec:gasSIMU}).  Spectroscopic
determinations of YSO binary component masses are still rare (e.g.,
{\em Daemgen et al.}, 2012, 2013; {\em Correia et al.}, 2013), and
estimates of mass ratios for young stars are mostly based on component
photometry, with the significant caveats that come from accretion
luminosity, differences in extinction of the components, and biases
towards detecting brighter companions. For young intermediate mass
stars in the Sco~OB2 association, {\em Kouwenhoven et al.} (2005)
could fit the mass ratio distribution with a declining function for
rising mass ratios (${f(q)}\sim{q^{-0.33}}$), revealing a clear
preference for low-$q$ systems. In contrast, low mass YSOs have a
gently rising distribution for rising mass ratios, which becomes
increasingly steep for VLM objects, showing a clear preference for
$q$$\sim$1 binaries ({\em Kraus et al.} 2011; {\em Kraus and
  Hillenbrand}, 2012), as do VLM binaries in the field (e.g., {\em
  Burgasser et al.}, 2007). It is noteworthy that this naturally
results from dynamical interactions in VLM triple systems ({\em
  Reipurth and Mikkola}, 2014).

\subsection{Eccentricities}
\label{sec:pmsECC}

The eccentricity of binaries in the solar neighborhood has been
studied by {\em Raghavan et al.} (2010), who finds an essentially flat
distribution from circular out to $e$$\sim$0.6 for binaries with
periods longer than $\sim$12~days (to avoid the effects of
circularization). Higher eccentricities are less common, but this may
be due to observational bias.
For VLM binaries, {\em Dupuy and Liu} (2011) found eccentricities with
a distribution very similar to the solar neighborhood.

Little is known about eccentricities of young binaries, except
for the $\sim$50 mostly short-period spectroscopic PMS binaries that
have been analyzed to date; {\em Melo et al.} (2001) found that
binaries with periods less than 7.5~days have already circularized
during pre-main sequence evolution, in agreement with theory ({\em
  Zahn and Bouchet}, 1989).

\section{\textbf{PRE-MAIN SEQUENCE ECLIPSING BINARIES}}
\label{sec:eb}

Accurate measurements of the basic physical properties -- masses,
radii, temperatures, metallicities -- of PMS stars and brown dwarfs are
essential to our understanding of the physics of star formation.
Dynamical masses and radii from eclipsing binaries (EBs) remain the
gold standard, and represent the fundamental testbed with which to
assess the performance of theoretical PMS evolution models.  In turn
these models are the basis for determining the basic properties of all
other young stars generally -- individual stellar masses and ages, mass
accretion rates -- and thus help to constrain key aspects of
star-forming regions, such as cluster star-formation histories and
initial mass functions.

\subsection{Performance of PMS Evolutionary Models}
\label{sec:ebMODELS}

The {\em PPV} volume included a summary of the properties of the 10 PMS
stars that are components of EBs known at that time ({\em Mathieu et
  al.}, 2007), and summarized the performance of four different sets
of PMS evolutionary tracks ({\em D'Antona and Mazzitelli}, 1997; {\em
  Baraffe et al.}, 1998; {\em Palla and Stahler}, 1999; {\em Siess et
  al.}, 2000) in predicting the dynamically measured masses of these
stars from their H-R Diagram positions.  
To summarize briefly the current status: {\em (1)} All of the
above models correctly predict the measured masses to $\sim$10\% above
1~M$_\odot$; {\em (2)} the models overall perform poorly below 1~M$_\odot$,
generally predicting masses larger than the observed masses by up to
100\%, and {\em (3)} the models of {\em Palla and Stahler} (1999) and {\em
  Siess et al.} (2000) are performing the best, predicting the
observed masses to 5\% on average although with a large scatter of
25\%.

There are now as of this writing 23 PMS stars that are components of
13 EBs, including two brown dwarfs in one EB ({\em
  Stassun et al.}, 2006, 2007).  An important development is the
emergence of new models -- the first in more than a decade -- with
physics attuned to PMS stars, namely the Pisa models ({\em Tognelli et
  al.}, 2011) and the Dartmouth models (e.g., {\em Dotter et al.},
2008; {\em Feiden and Chaboyer}, 2012). A full assessment of the latter
models against the sample of PMS EBs is underway ({\em Stassun et
  al.}, in prep.), but preliminary results are promising.  For
example, the dynamically measured masses are correctly predicted by
the Dartmouth models to $\sim$15\% over the range of masses 0.2--1.8
M$_\odot$.

The major review by {\em Torres et al.}\ (2010), while focused on
main-sequence EBs, highlights the importance of reliable
metallicities, temperatures, and (when possible) apsidal motions.
Among PMS EBs, metallicity determinations are not commonly reported,
but should in principle be determinable from the spectra used for the
radial-velocity measurements. Temperatures remain vexing because of
uncertainties over the spectral-type to temperature scale for PMS
stars, especially at very low masses.  Only recently was the first
apsidal motion for a very young EB reported (V578 Mon; {\em Garcia et
  al.}, 2011, 2013a).  As demonstrated by {\em Feiden and Dotter}
(2013), such apsidal motion measurements can provide particularly
stringent constraints on the models, specifically on the interior
structure evolution with age, a critically important physical
ingredient.

\begin{figure}[ht]
\includegraphics[width=\linewidth]{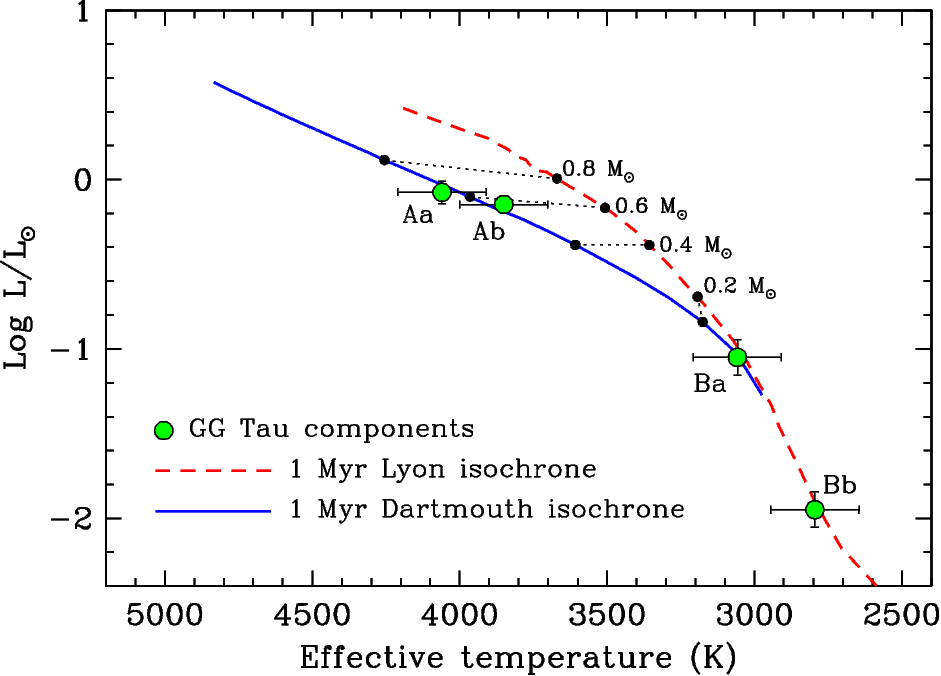}
\caption{\label{fig-ggtau}
\small 
Application of Dartmouth models to the quadruple PMS system GG Tau
({\em Torres et al.}\ 2013).
Previous generation models (here compared to the Lyon
models), which have been used to calibrate the PMS temperature scale, do not
perform as well.}
\end{figure}

Importantly, {\em Torres et al.} (2013) have used the quadruple PMS
system LkCa3 to perform a stringent test of various PMS evolutionary
models. They find clearly that the Dartmouth models perform best, and
moreover find that these models can fit another benchmark quadruple
system, GG Tau, whereas previous generation models cannot
(Fig.~\ref{fig-ggtau}).

\subsection{Impact of Activity on Temperatures, Radii, and Estimated
Masses of Young Stars}
\label{sec:ebACTIVITY}

Stars in short-period binaries are often chromospherically active, and
thus may suffer from activity-reduced temperatures and/or -inflated
radii, causing them to appear discrepant relative to standard model
isochrones (e.g., {\em Torres}, 2013).  In particular, such
activity-reduced temperatures can cause the derived stellar masses
to be underestimated by up to a factor of $\sim$2.

A particularly important case in point is 2M0535--05, the only known
EB comprising two brown dwarfs ({\em Stassun et al.}, 2006, 2007),
which proved enigmatic from the start. The system is a member of the
very young Orion Nebula Cluster, and thus the expectation is that both
components of the EB should have an age of $\sim$1 Myr. However, a
very peculiar feature of the system is a reversal of temperatures with
mass -- the higher mass brown dwarf is cooler than its lower-mass brown
dwarf companion -- making the higher mass brown dwarf appear younger
than the lower-mass companion and a factor of 2 lower in mass than its
true mass. 
{\em Reiners et al.} (2007) showed that the higher mass brown dwarf is
highly chromospherically active as measured by the luminosity of its
H$\alpha$ emission, whereas the lower mass brown dwarf is a factor of
10 less active and appears ``normal'' relative to the evolutionary
tracks.

Motivated by this peculiar but important system, {\em Stassun et al.}\
(2012) have used a sample of low-mass EBs to determine empirical
corrections to stellar temperatures and radii as a function of
chromospheric H$\alpha$ activity ({\em Morales et al.}, 2010).
Notably, these corrections indicate that the nature of the temperature
reduction and radius inflation is such that the bolometric
luminosity is roughly conserved. The {\em Stassun et al.}\ (2012)
relations are able to fully explain the anomalous temperature reversal
found in the 2M0535--05 brown-dwarf EB.

However, there is not as yet consensus on the underlying physical
cause of this effect. {\em Chabrier et al.}\ (2007) suggest that
surface spots and convection inhibited at the surface are the driver,
whereas {\em MacDonald and Mullan} (2009) suggest a global inhibition
of convection through strong fields threading the interiors of the
stars.  {\em Mohanty and Stassun} (2012) performed detailed
spectrocopic analysis of the eclipsing brown-dwarf EB 2M0535--05, the
results of which appear to disfavor the {\em Chabrier et al.}\ (2007)
hypothesis. However, questions remain as to the physical plausibility
of the magnitude of interior fields required in the {\em MacDonald and
  Mullan} (2009) hypothesis.

At the same time, the Dartmouth models also now incorporate the
effects of internal magnetic fields, which successfully accounts for
the effects of temperature reduction and radius inflation in a
physically self-consistent fashion ({\em Feiden and Chaboyer}, 2012).

\subsection{Impact of Triplicity on Properties of PMS Stars}
\label{sec:ebTRIPLES}

There is increasing evidence that the presence of third bodies in
young binaries can significantly alter the properties of the component
stars, either directly through tidal heating effects and/or indirectly
by impacting the accretion history of the system.  As an exemplar
case, {\em Stassun et al.}\ (2008) identified Par 1802 to be an
unusual PMS EB whose component stars have identical masses ($q=0.99$)
yet radii that differ by 7\%, temperatures that differ by 9\%, and
luminosities that differ by 60\%.  Thus the pair cannot be fit by any
standard PMS stellar evolution models under the usual assumption of
coevality for the component stars because the stars' highly unequal
luminosities cause them to appear highly non-coeval.  {\em Gomez
  Maqueo Chew et al.}\ (2012) used 15 yr of eclipse timing
measurements to reveal the presence of a wide tertiary component in
the system. Modeling the tidal heating on the EB pair arising from the
previous orbital evolution can explain the over-luminosity of the
primary eclipsing component, and moreover suggests a close three-body
(perhaps exchange) interaction in the past.

Relatedly, recent theoretical work has suggested that accretion
history (e.g., FU Ori outbursts, differential accretion in
proto-binaries) can alter the PMS mass-radius relationship (e.g., {\em
  Simon and Obbie}, 2009; {\em Baraffe and Chabrier}, 2010).
Consequently, new generation PMS evolution models are seeking to
simulate these effects. For example, the new Pisa models are being
further developed to include thin-disk accretion episodes during the
early PMS phase.

As suggested by the example of Par 1802 above, PMS EBs provide a
unique opportunity to assess the frequency of higher-order multiples
among close binaries, because of the high quality and multi-faceted
ways in which these benchmark systems are studied.  Among the sample
of 11 PMS EB systems that have detailed EB solutions published as of
this writing (i.e., excluding PMS EBs with preliminary reports such as
the 6 systems announced in {\em Morales-Calderon et al.}, 2012) and
that have stellar mass components (i.e., excluding the double
brown-dwarf EB 2M0535--05), 6 are now known to include a third body.
This preliminary census implies a very high ratio of triples to
binaries, consistent with the view that tertiaries may be critical to
the formation of tight pairs.

\section{\textbf{GAS IN BINARIES AND MULTIPLE SYSTEMS}}
\label{sec:gas}

\subsection{Observations of Circumbinary Structures}
\label{sec:gasCIRCUM}

Circumbinary disks play an important role in shaping binary orbital
properties: mass flow from the disk affects the ultimate binary mass
ratio while the flow of angular momentum from binary to disk drives
changes in the binary period and eccentricity. The observational study
of circumbinary accretion flows is, however, challenging: massive
circumbinary disks are rare amongst binaries with separations in the
range of a few to $100$ AU ({\it Jensen et al.}, 1996; {\it Harris et
  al.}, 2012), which constitute the bulk of the pre-main sequence
binary population. To date, only a handful of circumbinary disks have
been imaged directly (see {\it Hioki et al.}, 2009) and here the
limitations of coronagraphic imaging do not allow the study of the
structures -- critical to the binary's evolution -- that link the disk
to the binary.  Circumbinary disks are considerably more abundant
around the closest binaries (a few AU or less); on the main sequence
such binaries are -- unlike wider pairs -- preferentially associated
with circumbinary {\it debris} disks ({\it Trilling et al.}, 2007) and
are in the regime where {\it Kepler} has recently revealed a number of
circumbinary planets (e.g., {\it Doyle et al.}, 2011; {\it Welsh et
  al.}, 2012).  The reason for the higher incidence of massive
circumbinary disks in close pairs is unclear -- i.e.,  whether it
reflects the initial configuration at formation or whether such disks
drive binaries to small orbital separations.  Alternatively, this
association may be a matter of disk survival: {\it Alexander} (2012)
has argued that disks around close binaries should be long lived,
since viscous draining is impeded by the binary's tidal barrier while
the gas is too tightly bound to be readily photo-evaporated.

Interferometric studies are just beginning to probe the dust
morphology in these systems ({\it Boden et al.}, 2009; {\it Garcia et
  al.}, 2013b) and so the bulk of our knowledge derives from time
domain studies.  For example, in eccentric binaries, periodic optical
and X-ray variations have been attributed to a dynamically modulated
accretion flow (e.g., {\it Mathieu et al.}, 1997 in DQ Tau; {\it Gomez
  da Castro et al.}, 2013 in AK~Sco), although optical variability
also accompanies synchotron flares at mm wavelengths, which can be
understood as reconnection events when the two stellar magnetospheres
interact at periastron (e.g., {\it Salter et al.}, 2010).  {\it
  Muzerolle et al.}  (2013) have recently interpreted large scale
periodic variations in a {\it protostellar} source as deriving from a
binary-modulated pulsed accretion flow.  Variations in the observer's
viewing angle also modulate line emission in low eccentricity
binaries: for example in V4046 Sgr, hydrodynamical modeling ({\it de
  Val-Borro et al.}, 2011) reproduces the periodic changes in the
wings of the Balmer lines observed by {\it Stempels and Gahm} (2004).
In CS~Cha, the binary's variable illumination of dusty accretion
streams has been invoked to explain its periodic infrared variability
({\it Nagel et al.}, 2012); it is, however, notable that the spectral
energy distribution of CS~Cha implies that the inner edge of the
optically thick circumbinary disk is at about $10 \times$ the binary
orbital separation ({\it Espaillat et al.}, 2011), which is several
times larger than what is expected from dynamical truncation by the
binary.  This finding exemplifies the difficulty of connecting models
and observations, since the dust emission is apparently not merely
being shaped by the response of the gas to the binary potential.

\subsection{Simulations of Circumbinary Disks}
\label{sec:gasSIMU}

While observed circumbinary disks are generally low in mass during the
Classical T Tauri phase, this was almost certainly not the case at
earlier evolutionary phases. In hydrodynamic collapse simulations,
proto-binaries are surrounded by circumbinary disks formed from higher
angular momentum material in the natal core, and the interaction
between the disk and the binary is key to shaping the system's
ultimate orbital elements ({\em a, e, q}). A complete theory of
binary formation should require not only the creation of the
protobinary fragments but should contain a clear prescription for the
evolution of these quantities as a function of the properties of the
circumbinary disk.

Unfortunately this goal remains elusive, despite comprehensive (SPH)
studies devoted to this problem (e.g., {\em Bate and Bonnell}, 1997;
{\em Bate}, 2000).  Qualitative features of these studies (especially
the preferential accretion of gas onto the secondary and hence the
increase of the binary mass ratio) were challenged by {\it Ochi et
  al.} (2005) and {\it Hanawa et al.} (2010) whose AMR simulations
were morphologically distinct and involved a preferential accretion of
gas onto the primary (thus driving $q= M_2/M_1$ downwards). It now
seems likely that these differences arose from the different
parameters of the latter studies (i.e., warm, two dimensional flows)
rather than from a code difference: nevertheless there are no fully
converged simulations of circumbinary accretion that have been run to
a steady state and this probably explains the variety of results
reported in the literature with regard to the sign and magnitude of
effects associated with circumbinary accretion (see also {\it Fateeva
  et al.}, 2011; {\it de Val Borro et al.}, 2011). This raises a
cautionary note with regard to the fidelity of cluster scale
simulations in modeling this process, since disks in such simulations
are always relatively poorly resolved.

If there is still no clear consensus in the purely hydrodynamical
case, the situation becomes still more complicated when magnetic
fields are involved. This is illustrated by two recent studies. {\it
  Zhao and Li} (2013) modeled accretion onto a `seed binary' placed
within a moderately magnetised core and found that severe magnetic
braking of the accreting gas has two notable effects: the binary
shrinks to small separations, while the low angular momentum of the
braked gas ensures that the flow is predominantly to the primary, thus
lowering the binary mass ratio.  In another study, {\it Shi et al.}
(2012) conducted the first simulation of binary/circumbinary disk
interaction which -- rather than adopting a parameterized
`$\alpha$-type' viscosity in the disk ({\it Shakura and Sunyaev},
1973) as in most previous works (e.g., {\it Artymowicz and Lubow},
1994; 
{\it MacFadyen and Milosavljevic},
2008; {\it Cuadra et al.} 2009; {\it Hanawa et al.}, 2010) -- instead
simulated the self-consistent angular momentum transfer associated
with the development of magneto-hydrodynamic turbulence in the disk.
The simulation considered the limit of ideal MHD and is therefore not
applicable to `dead zones' of low ionization ({\it Bai}, 2011; {\it
  Mohanty et al.}, 2013): in practice this limits it to radii within
$\sim 0.5$ AU or beyond $\sim 10$ AU.

It is found that the effective efficiency of angular momentum
transport (e.g., as parameterized by the {\it Shakura and Sunyaev}
$\alpha$ parameter) is about an order of magnitude higher in the
accretion streams that link the binary to the disk than in the body of
the disk, and this results in a much more vigorous flow through the
accretion streams than in previous simulations that do not treat the
development of magneto-turbulent stresses self-consistently (see {\it
  MacFadyen and Milosavljevic}, 2008). Indeed in the {\em Shi et al.}
(2012) simulations the flow through the accretion streams is $\sim 30
\%$ of the flow through the outer disk.  In such a situation the net
evolution of the binary is governed by two nearly cancelling terms
(the spin-up effect of accretion and the spin-down torque associated
with the non-axisymmetric disk/accretion streams) and is thus very
sensitive to numerical inaccuracies/uncertainties in the disk
thermodynamics.  So although this simulation is undoubtedly more
realistic than previous calculations, it raises awkward issues:
apparently the derivation of a simple relationship between
circumbinary disk properties and associated orbital evolution may be
more elusive than ever.

\subsection{Disk Lifetimes in Binaries}
\label{sec:gasLIFETIME}

Since the review of this subject by {\it Monin et al.} (2007) in PPV,
a number of studies have charted the relative lifetimes of disks in
binaries compared with single stars, and studied the relative
lifetimes of the primaries' and secondaries' disks.  Early studies in
this area (e.g., {\it Prato and Simon}, 1997) had argued that
circumstellar disks must be replenished from a circumbinary reservoir
during the Classical T Tauri phase, a requirement that was puzzling
given the observed lack of circumbinary material in all but the
closest binaries. This conclusion was based on {\em a)} the fact that
disks in binaries were not apparently shorter lived than disks in
single stars and {\em b)} the scarcity of `mixed pairs' (i.e., those
with only one disk). Re-supply of circumstellar disks would extend
their lifetimes and coordinate the disappearance of the disks, since
otherwise -- in isolation -- the secondary's disk would drain first on
account of its smaller tidal truncation radius and shorter viscous
timescale ({\it Armitage et al.}, 1999).

However, recent studies have undermined the observational basis for
these arguments. {\it Cieza et al.} (2009) and {\it Kraus and
  Hillenbrand} (2009) demonstrated that the lifetime of disks in close
binaries is indeed reduced compared with single stars or wide pairs,
concluding that incompleteness of the census of close binaries, the
use of unresolved disk indicators and projection effects had all
previously masked this correlation in smaller samples (see also {\it
  Kraus et al.}, 2012; {\it Daemgen et al.}, 2013). Moreover, the census
of binaries for which spectral diagnostics have been measured for each
component has been augmented by {\it Daemgen et al.} (2012, 2013) in the
ONC and Chamaeleon I. These new results have reinforced the suggestion
of {\it Monin et al.} (2007), that the early conclusions about the absence of
mixed pairs was skewed by results from Taurus which are not borne out
in other regions. {\it Kneller and Clarke} (2014) argue
that the observed incidence of disks in binaries as a function of $q$
and separation is compatible with clearing by combined viscous
draining and X-ray photo-evaporation. Such models predict a strong
tendency for the secondary's disk to disappear before the primary's
for binaries closer than $100$ AU while predicting that in wider mixed
pairs, disks are equally likely to exist around the secondary and
primary components: this latter prediction needs to be tested
observationally in larger samples.

\subsection{Eclipses by Disks}
\label{sec:ebDISKS}

Disks may cause eclipses of their central stars, and these events
present rare but valuable opportunities to study the detailed
structure of disks during the planet forming era.  The best studied
example is KH15D (e.g., {\em Herbst et al.}, 2010).  KH15D is a binary
that is occulted by a circumbinary screen of material that moves
slowly across the binary components, occulting them in turn.  Modeling
of the screen suggests its origin to be a precessing, warped
circumbinary ring of material several AU from the tight binary. The
obscuring ring has very sharp edges -- it is well modeled as a knife
edge -- indicating a high degree of coherence to the material despite
the dynamics of the system. 

\begin{figure}[t]
 \epsscale{0.9}
\plotone{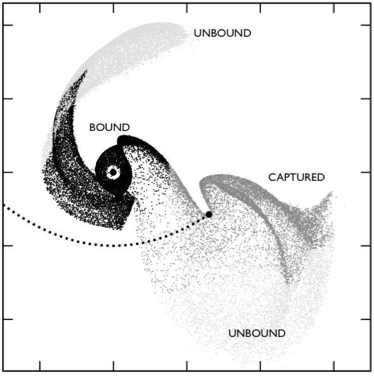}
  \caption{\small Severe disk disturbances occur during close periastron 
passages. This simulation shows an encounter between two solar-type stars, 
one with and one without a disk; black dots show material that remains bound, 
grey dots show material captured by the intruder, and light grey dots show 
unbound material. The box is 500 $\times$ 500~AU.
Courtesy S. Pfalzner and M. Steinhausen.
\label{fig:pfalzner}
}
 \end{figure}

RW Aur is a newly discovered exemplar of this class ({\em Rodriguez et
  al.}, 2013). In this case, light curve observations by the KELT
exoplanet transit survey ({\em Pepper et al.}, 2007) witnessed a
sudden dramatic eclipse of the star with a depth of $\sim$2 mag and
lasting approximately 180 days.  Archival photometric observations can
rule out a similarly long and deep eclipse over the past 100 years.
This singular event is interpreted and modeled as an occultation of
the primary star (RW Aur A) by the long tidal arm observed by {\em
  Cabrit et al.}\ (2006) resulting from tidal disruption of its
circumstellar disk by the recent fly-by of RW Aur B (see
Figure~\ref{fig:pfalzner}).  RW Aur B may itself be a tight binary,
making the RW Aur system a triple ({\em Ghez et al.}, 1993).  The
eclipse observations indicate a knife-edge structure to the occulting
feature, consistent with the dynamical simulations of {\em Clarke and
  Pringle} (1993), which demonstrate a high degree of coherence in the
tidal arm persisting long after the fly-by (see
Section~\ref{sec:retention}).

\subsection{Alignment of Orbital Planes in Young Binaries}
\label{sec:alignment}

A number of systems show evidence that they have undergone dynamical
events that have perturbed the orientation of the binary and/or its
circumstellar disks. For example, {\it Bisikalo et al.} (2012)
presents evidence that the disk around RW Aur A (discussed above) is
{\it counter-rotating} with respect to the binary orbital motion.
Similarly, {\it Albrecht et al.} (2009) used Rossiter-McLaughlin
measurements during the eclipse of the massive binary DI Her to show
that the projected spin of one of its B-type components was highly
misaligned (72$^{\circ}$) with respect to the binary orbital plane.
There is no unique explanation for such systems. One idea is that,
whereas the spin of the stars reflects that of the local gas reservoir
(material that collapsed first), the spin direction of circumbinary
structures (or the orbital plane of binary systems) may inherit a
different direction from a larger region within the turbulent medium,
because of chaotic changes in the mean angular momentum vector of
accreting material (this effect is significant in the whole cluster
simulations of {\it Bate} 2009a, where misaligned systems are
common). Alternatively, dynamical interactions (for example an
exchange interaction within a non-hierarchical system) can play a
similar role, although this again requires that the natal gas contains
a range of spin directions. On the other hand, the Kozai-Lidov ({\it
  Kozai}, 1962; {\it Lidov}, 1962) mechanism within triple systems can
induce spin-orbit misalignment even in the absence of exchange
interactions: {\it Fabrycky and Tremaine} (2007) suggested that while
Kozai-Lidov induced oscillations in eccentricity and inclination can
deliver companions to small peri-center distances, tidal dissipation
could allow such systems to free themselves from the Kozai-Lidov
regime, trapping them in a state where their spins are decoupled from
their orbital inclination. Triple companions, however, have not been
detected to date in either of these systems. Other recent measurements
of misalignment of orbital planes within pre-main sequence binaries
include KH15D ({\it Capelo et al.}, 2012) and FS Tau ({\it Hioki et
  al.}, 2011), while circumbinary {\it debris} disks present a mixed
picture with respect to the alignment of orbital and circumbinary disk
planes ({\it Kennedy et al.}, 2012a,b).

{\it Facchini and Lodato} (2013) and {\it Foucart and Lai} (2013) have
recently presented analytic and numerical calculations of the
evolution of the warp and twist of a disk that is initially misaligned
with the binary orbit.  {\it Foucart and Lai} showed that the
back-reaction on the binary orbit re-aligns the system on a timescale
that is short compared with that required for the binary to accrete
significant mass from the circumbinary disk. They therefore argued
that close binaries (which gain significant mass from the circumbinary
disk) should become aligned with their disks during the pre-main
sequence period, thus explaining the surprising abundance of
(necessarily aligned) planets in circumbinary orbit in the Kepler
sample (e.g., {\it Doyle et al.}, 2011; {\it Welsh et al.}, 2012).

\subsection{Retention of Disks in Dynamical Encounters}
\label{sec:retention}

Dynamical interactions within multiple systems result in the pruning
of circumstellar disks, leading {\it Reipurth and Clarke} (2001) to
argue that disk {\it size} may provide a diagnostic of an object's
previous history of close encounters in a few-body system. The
influence of stellar fly-by's on disk structure was first examined by
{\it Clarke and Pringle} (1993), while {\it Hall} (1997) reconstructed
disk surface profiles post-encounter from ballistic calculations
through the assumption that bound particles should re-circularize
while retaining their individual angular momenta. The SPH calculations
of {\it Pfalzner et al.}  (2005) (see Figure~\ref{fig:pfalzner})
showed that this is a reasonable approximation, and more recently {\it
  Umbreit et al.}  (2011) have applied the same procedure to stars
undergoing close encounters within triple systems. This study (which
started from non-hierarchical co-planar triples with co-planar disks)
showed that the reconstructed density profiles show a boosted power
law profile in the inner disk and an exponential cut-off at a radius
of a few tenths of the minimum encounter distance.  It is found that
disk stripping during triple decays is qualitatively very similar and
only slightly stronger than that occurring during two-body fly-by's.

\subsection{Planetary Systems in Multiple Stellar Systems}
\label{sec:planets}

Around $7\%$ of currently detected planets are in binary systems, most
of which are located in circum{\it stellar} orbits in wide systems.
Two categories of system have attracted considerable recent interest,
i.e.,  the circumbinary (P-type) planets around close (sub-AU)
separation binaries discovered by Kepler ({\it Doyle et al.}, 2011;
{\it Welsh et al.}, 2012) and the
circumstellar (S-type) planets discovered in relatively close ($a <
20$ AU) pairs ({\it Chauvin et al.}, 2011; {\it Dumusque et al.},
2012).  Planetesimal accumulation in binary environments faces a well
known problem ({\it Thebault et al.}, 2008) on account of the pumping
of the planetesimal velocity dispersion by gravitational perturbations
from the binary; a high velocity dispersion implies destructive
collisions ({\it Leinhardt and Stewart}, 2012) and thus limits the
possibilities for planetesimal growth.  Some suggested solutions to
this problem fall into the category of simply forming planets in a
more benign dynamical regime (i.e., further from the perturber) and
then invoking migration -- of either the planet or the binary
companion -- to achieve the observed planet/binary architecture: see
{\it Payne et al.} (2009), {\it Thebault et al.}  (2009).
Alternatively, {\it Xie et al.} (2010) have explored the effect of a
modest inclination between the binary and disk plane. In terms of {\it
  in situ} formation models, gas drag has been invoked as a mechanism
for enforcing apsidal alignment of perturbed planetesimal orbits: this
produces local velocity coherence in objects of a given size. However,
since gas drag is a size-dependent phenomenon, this does not prevent
destructive collisions in a planetesimal population with a realistic
size distribution and is thus unlikely to solve the problem ({\it
  Thebault et al.}, 2006, 2008).

Recent works on this topic concentrate on the effect of the disk's
gravitational field upon the growth of planetesimal velocity
dispersion.  {\it Rafikov} (2013) argued that {\it if} the disk is
approximately axisymmetric, then its gravitional influence induces {\it
  size independent} apsidal precession, which acts to reduce the
eccentricity excitation by the companion.  However, the simulations of
{\it Marzari et al.} (2013) show that the planet induces a strong
eccentricity in the disk, and that gravitational coupling with the
eccentric disk actually amplifies the stirring of the planetesimal
population.  In effect, therefore, these studies come to qualitatively
opposite conclusions as to whether binarity is a major obstacle to
planet formation. These divergent conclusions essentially hinge on the
axisymmetry of the gas disk in the region of interest; further hybrid
hydrodynamical/N-body modeling is required in order to delineate the
areas of parameter space in which planetesimal growth is possible.

Finally, although planets are known in stellar triple systems (e.g.,
{\em Bechter et al.}, 2013), the issue of planet formation in higher
order multiples is unexplored (although several works have examined
the stability boundaries of particle disks within triple systems,
e.g., {\it Verrier and Evans}, 2007; {\it Domingos et al.}, 2012).
While the presence of three bodies in general restricts the
stability regions available, there are certain configurations where a
third body can stabilize particle orbits.  In particular, whereas
particles in a circumstellar disk with an inclined companion can be
subject to Kozai-Lidov instability, this can be suppressed if the
central object within the disk is itself a binary, since in this case
the binary induces nodal libration which stabilizes the particles
({\it Verrier and Evans} 2009). Such studies will be important in
interpreting the statistics of debris disks within multiple star
systems.

\section{\textbf{BINARIES IN CLUSTERS}}
\label{sec:cluster}

We have seen (see Section~\ref{sec:dynamics}) that multiple systems
can change due to internal (secular) interactions.  But dynamical
interactions with other systems can play an extremely important role
in altering multiple systems (changing their orbital parameters, or
even destroying them).  In the relatively dense environments of star
forming regions or star clusters the initial multiple population can
be very significantly altered.  This means that any multiple
population we observe is almost certainly not what formed, and
different populations can evolve in (very) different ways depending on
the environment.  Therefore, to extrapolate back to formation from any
observations, we must fold-in the (possibly very complex) dynamical
evolution (called `inverse population synthesis' by {\em Kroupa},
1995).

\subsection{External dynamical interactions}
\label{sec:clusterEXT}

The dynamical destruction of binaries was first studied in detail by
{\em Heggie} (1975) and {\em Hills} (1975; see also {\em Hills},
1990).  They placed binaries into two broad categories: `hard' and
`soft'.  Hard binaries are those which are so tightly bound that
encounters are extremely unlikely to alter their properties (to
destroy them, or even to change their orbital parameters much).  Soft
binaries are those that are so weakly bound, that they are almost
certain to be destroyed by any encounter.  Generally, if the kinetic
energy of a perturber is greater than the binding energy of the
binary, then the binary is destroyed.  An alternative way of looking
at this is that a binary is hard if its orbital speed is greater than
the speed of an encounter ({\em Hills}, 1990).  Investigation of the
dynamics of encounters leads to the Heggie-Hills law, that states that
hard binaries get harder (encounters typically take energy from the
binary), and soft binaries get softer (encounters typically give
energy to the binary).

The hard-soft boundary is basically set by the velocity dispersion of
the perturbing stars, and the mass of the binary.  The faster that
perturbers are moving, the closer the hard-soft boundary, and the more
massive the binary is, the more difficult it is to destroy.  It might be
expected that binary destruction will depend on the mass ratio,
however simulations show that destructive encounter energies are almost always
significantly greater than the binding energy, and so destruction does
not depend on the mass ratio ({\em Parker and Reggiani}, 2013).

However, the `hardness' of a binary is not the only thing that decides
if a binary will survive.  To destroy even a soft binary an
encounter is required, therefore the encounter rate 
is crucial.  In the field, many formally soft binaries survive for
significant amounts of time, because the encounter rate is very low.

Therefore, the survival of a binary depends on {\em (a)} the energy of
the binary, {\em (b)} the energy of encounters, and {\em (c)} the
frequency of encounters.  The more massive and dense a star-forming
region or star cluster is, the more frequent and energetic encounters
will be, and so binary destruction/alteration should be more
efficient.

In any given environment, hard binaries should survive, and soft 
binaries will almost certainly be destroyed.  The most interesting
binaries, however, are often `intermediates' between hard and soft, which may
or may not survive depending on the exact details of their dynamical
histories (e.g., {\em Parker and Goodwin}, 2012).

Let us take an `average' binary system of component masses $m$, where
$m \sim 0.4~$M$_\odot$ is the average mass of a star.  Let us put this
binary in a virialized cluster of $N$ stars of total mass $M=Nm$, and
radius $R$.  The hard-soft boundary, $a_{\rm hs}$ will be at
approximately 
\begin{equation}
a_{\rm hs} \sim 10^5 \left( \frac{R}{\rm pc} \right) \left(
\frac{1}{N} \right) \,\,\,\,{\rm AU}
\end{equation}
(see {\em Parker and Goodwin}, 2012).
Numerical experiments show that a safe value for a hard binary that
will almost certainly survive is about $a_{\rm hs}/4$.  For clusters
like the Orion Nebula Cluster (ONC; $N \sim 10^3$, $R \sim 1$~pc)
$a_{\rm hs} \sim 100$~AU.  For relatively nearby regions, the ONC is
very massive and dense, and so in local regions we tend not to expect
processing of binaries with $a < 100$~AU.  This means that the $a <
100$~AU population of binaries is `pristine' (i.e., unprocessed),
whilst $a > 100$~AU binaries may (or may not) have been processed
({\em Goodwin}, 2010; {\em King et al.}, 2012b).

It is important to remember, however, that it is not necessarily 
the {\em current} density of a region that is important in assessing
the possible impact of binary processing.  Rather, it is the (usually
unknown) density history of the region.  

The values of $N \sim 10^3$ and $R \sim 1$~pc used above for the ONC
are the present-day values, and the calculated hard-soft boundary of $a
\sim 100$~AU is the current safe hard-soft boundary.  If the ONC was
much denser in the past (as has been argued, see {\em Scally et al.},
2005; {\em Parker et al.}, 2009), then the hard-soft boundary in the
past could have been much smaller.  If a region spends at least a
crossing time in a dense state, then it is that dense state that
imposes itself in binary destruction ({\em Parker et al.}, 2009).
This could be very important if regions undergo expansion due to gas
expulsion (e.g., {\em Marks and Kroupa}, 2011, 2012), or process
binaries in short-lived substructures ({\em Kroupa et al.}, 2003; {\em
  Parker et al.}, 2011).

\subsection{Binary Formation through Encounters}
\label{sec:clusterENCOUNTER}

Dynamics are usually associated with binary destruction rather than
formation.  But hard binaries can be formed by three-body encounters (with
the third body carrying away energy).  The rate of binary creation per
unit volume, $\dot{N_{\rm b}}$, depends on the stellar masses ($m$),
velocity dispersion ($\sigma$), and number density of stars ($n$)
\begin{equation}
\dot{N_{\rm b}} = 0.75 \frac{G^5 m^5 n^3}{\sigma^9}
\end{equation}
({\em Goodman and Hut}, 1993).  In the Galactic field this number is
essentially zero ($\sim 10^{-21}$~pc$^{-3}$~Gyr$^{-1}$).  However, in
dense star-forming regions and clusters the rate may be significant,
especially for higher-mass stars.  Simulations show that initially
single massive stars can pair-up in hard binaries, and can form
complex higher-order systems similar to the Trapezium ({\em Allison and
  Goodwin}, 2011).  This is due to the very strong dependency of
$\dot{N_{\rm b}}$ on the higher $m$ and $n$, and the
lower $\sigma$ in clusters (which can make 30 orders of magnitude
difference).  

{\em Kouwenhoven et al.} (2010) and {\em Moeckel and Bate} (2010)
independently found that dissolving dense regions can also form very
wide binaries by `chance', when two stars leaving the region find
themselves bound once outside of the region.  Similarly, {\em Moeckel
  and Clarke} (2011) find that dense regions constantly form soft
binaries.  While the region remains dense, these binaries are destroyed
as fast as they are made.  However, when the region dissolves into the
field they can be `frozen in' at lower densities and survive.  On
average, one region produces one wide binary with a median separation
of about $10^4$~AU, almost independently of the number of stars in that
region ({\em Kouwenhoven et al.}, 2010). 
Since the stars are paired randomly, it is quite possible
for the wide binaries to be made of one or two hard binaries (making
triple or quadruple systems).  The mass ratio distribution of wide
binaries would be expected to be randomly paired from the IMF.
This process acts independently of the wide binaries that form through
the unfolding of triple systems, as discussed in Section~\ref{sec:dynamics}.

\subsection{Observations of Young Multiples in Clusters}
\label{sec:clusterOBS}

It is only in nearby star-forming regions that we can examine in any
detail the (especially low-mass) binary properties.  Locally, young
star-forming regions cover a wide range of densities from a few
stars~pc$^{-3}$ (e.g., Taurus) to a few thousand stars~pc$^{-3}$ (e.g.,
the ONC; see {\em King et al.}, 2012a,b).  These are often -- rather
arbitrarily -- divided by density into low-density `associations', and
high-density `clusters'.  More formally, `clusters' are often thought
of as bound objects, or objects at least a few crossing times old
(e.g., {\em Gieles and Portegies Zwart}, 2011).  We will take the {\em
  Gieles and Portegies Zwart} (2011) definition of a cluster as
dynamically old systems, since these are systems which we might expect to
have significantly processed their multiple populations.  Locally,
this probably safely includes the ONC and IC~348 as `clusters' for
which we have some detailed information on the stellar multiplicity.

It is worth noting that the level of processing of binaries will not
simply depend on mass, but rather on the dynamical age of a system.
For example, {\em Becker et al.} (2013) suggest that the binary
properties (and unusual IMF) of the low-mass `cluster' $\eta$ Cha
could be explained by an initially very high density and rapid
dynamical evolution.

Observations of the ONC and IC~348 show a lower binary frequency than
associations (e.g., {\em K\"ohler et al.}, 2006; {\em Reipurth et al.},
2007 for the ONC; {\em Duch\^ene et al.}, 1999 for IC~348).  The ONC
is also found to have an almost complete lack of wide ($>1000$~AU)
systems ({\em Scally et al.}, 1999).  

{\em King et al.} (2012a,b) collated binary statistics for 7 young regions
and attempted to correct for the different selection effects and
produce directly comparable samples.  Only in the range 62--620~AU is
it possible to compare regions as diverse as Taurus (with an average
density of $<10$ stars pc$^{-3}$) to the ONC (around $5000$ stars
pc$^{-3}$).  In this separation range, the binary fraction of Taurus
is around $21 \pm 5$\%, compared to around 10\% in regions with
densities greater than a few 100 stars pc$^{-3}$ (Cha I, Ophiuchus, IC~348,
and the ONC).  The Solar field values in the same range are roughly
10\%.

Given the densities of the ONC, it is almost impossible to
imagine a scenario in which we are observing the birth population.
The binaries we observe have separations of 62--620~AU, almost all above
the hard-soft boundary in the ONC.  Taking a size for the ONC of 1~pc,
a density of $5000$ stars pc$^{-3}$, and a velocity dispersion of 2 km
s$^{-1}$, the typical encounter timescale at 1000~AU is about a Myr --
roughly the age of the ONC.  

It is often stated that clusters have a field-like binary
distribution, however this is somewhat misleading.  Binary studies of
the ONC and IC348, the only dense clusters analyzed so far, are of a
limited range of around 50--700~AU and in this range they have a
similar binary fraction to the field.  However, we have no information
on smaller separations, and they are certainly not field-like at large
separations, where there is an almost complete lack of systems.

{\em Reipurth et al.} (2007) find a significant (factor of 2--3) 
difference between the ratio of wide (200--620~AU) to close 
(62--200~AU) binaries between the inner~pc of the ONC and outside of
this.  This could suggest a difference in dynamical age, and hence the
degree of processing, between the inner and outer regions of the ONC
({\em Parker et al.}, 2009).

Interestingly, {\em King et al.} (2012b) find that whilst the binary
frequency in the ONC is significantly lower than in associations, the
binary separation distribution looks remarkably similar.  Such
distributions in the 62--620~AU range are always approximately
log-flat in all regions and show no statistically significant
differences.  Taurus has twice as many binaries as the ONC in the same
separation range, but the distribution of binary separations is the
same.

This is worth remarking on, because it is very unexpected.  A
reasonable assumption would be that associations and clusters form the
same primordial population, but that clusters are much more efficient at
processing that population.  The field is then the sum of relatively
unprocessed binaries from associations, and relatively highly
processed binaries from clusters (e.g., {\em Kroupa}, 1995; {\em Marks
  and Kroupa}, 2011, 2012).  But processing is separation-dependent, and
wider binaries should be processed more efficiently than closer
binaries.  Therefore, if we take initially the same binary frequency
and separation distribution in the 62--620~AU range in both 
associations and clusters, we would expect {\em (a)} a
lower final binary frequency in clusters (which we see), and {\em (b)}
fewer wider binaries in clusters than associations (which we do not
see).  Note that by wide, we do not mean $>1000$~AU, which are missing
in the ONC ({\em Scally et al.}, 1999), but rather fewer, say,
200-620~AU binaries than 62--200~AU binaries.  

That the separation distributions in low-density and high-density
environments is the same could suggest that high-density regions
somehow over-produce slightly wider systems, which are then
preferentially destroyed to fortuitously produce the same final
separation distribution.  This would seem rather odd ({\em King et
  al.}, 2012b).

The 62--620~AU range of binaries for which we have
observations in the ONC are mostly (rather frustratingly) intermediate
binaries whose processing depends on the details of their dynamical
histories. {\em Parker and Goodwin} (2012) show that in ONC-like
systems the tendency is to preferentially destroy wider systems, but
small-$N$ statistics means that some clusters can produce separation
distributions in the observed range that sometimes retain the initial
shape.  So maybe the separation distribution in the ONC is
statistically slightly unusual?  However, the difference between the
inner and outer ratio of wide (200--620~AU) to close 
(62--200~AU) binaries observed by {\em Reipurth et al.} (2007)
suggests that the inner regions of the ONC have been efficient at
processing the wider binaries.

In summary, in clusters we expect significant binary destruction.
However, interpreting observations of binaries in clusters is
difficult.  This is due to the lack of nearby clusters, and the
limited range of binary separations that are observable.  But in the
binary populations of clusters should be clues to the formation and
assembly of clusters, and differences between star formation in
different environments.

\section{\textbf{THE  MULTIPLICITY OF MASSIVE STARS}} 
\label{sec:massive}

We here define massive stars to be OB stars on the main sequence,
above $\sim$ 10 M$_\odot$ (about B2V) capable of ionizing atomic
hydrogen, with the dividing line between O and B stars around 16
M$_\odot$ (about B0V, {\em Martins et al.}, 2005).  Massive stars
occur mostly in young clusters and associations, but to a small degree
also in the field and as runaway stars.  There are some 370 O-stars
known in the Galactic O Star Catalog ({\em Maiz-Apellaniz et al.},
2004; {\em Sota et al.}, 2008), with 272 located in young clusters and
associations, 56 in the field, and 42 classified as runaway stars.

\subsection{Recent Observational Progress}

A comprehensive review of the multiplicity of massive stars was given
by {\em Zinnecker and Yorke} (2007), emphasizing the difference in
multiplicity between high- and low-mass stars and its implication for
their different origins.  In the meantime, {\em Mason et al.} (2009)
in a statistical analysis summarized the multiplicity of massive stars
based on the Galactic O-star catalog (see above), both for visual and
for spectroscopic multiple systems.  {\em Chini et al.} (2012), in a
vast spectroscopic study, presented evidence for a nearly 100\% binary
frequency among the most massive stars, dropping substantially for
later-type B-stars, thus confirming the mass dependence of the
multiplicity.  At the same time, {\em Sana et al.} (2012) for the
first time derived the distributions of orbital periods and mass
ratios for an unbiased sample of some 70 O-stars based on a
multi-epoch, spectroscopic monitoring effort. Three important results
emerged: {\em (i)} the mass-ratio distribution is nearly flat with no
statistically significant peak at $q$=1 (identical twins); {\em (ii)}
the distribution of orbital periods peaks at very short periods (3-5
days) and declines towards longer periods; and {\em (iii)} a large
fraction ($>$70\%) of massive binaries are so close that the
components will be interacting in the course of their lifetime, thus
affecting the statistics of WR-stars, X-ray binaries, and supernovae,
and of these one third will actually merge ({\em Sana et al.}, 2012).

In yet another recent study, based on the VLT-FLAMES Tarantula Survey,
{\em Sana et al.} (2013) probed the spectroscopic binary fraction of
360 massive stars in the 30 Doradus starburst region in the Large
Magellanic Cloud.  They discovered that at least 40\% of the massive
stars in the region are spectroscopic binaries (both single and double
lined).  The unmistakable conclusion of all these studies is that the
processes that form massive stars strongly favor the production of
(mostly tight) binary and multiple systems.

Detailed studies of the multiplicity and orbital parameters of massive
stars in young clusters (NGC 6231, NGC 6611; {\em Sana et al.}, 2008,
2009) and OB associations (Cyg OB2, {\em Kiminki and Kobulnicky},
2012) have also been published, in an effort to find correlations with
cluster properties and statistical differences between cluster and
``field'' stars.  None were found (see the review by {\em Sana and
  Evans}, 2011).  A contentious issue is the multiplicity among {\em
  bona fide} runaway O-stars, which was believed to be low ({\em Gies
  and Bolton}, 1986; {\em Mason et al.}, 2009), but following the new
results of {\em Chini et al.}  (2012), it seems to be very high
(75\%).  In the case of runaway O-stars, it may eventually be useful
to discriminate between high-velocity runaway stars ($>$40~km/s),
presumably originating from supernovae explosions in binary systems
({\em Blaauw}, 1961), and slow runaways (``walk-aways'', $<$10~km/s,
which are harder to identify) whose origin is likely due to dynamical
ejection from dense young clusters ({\em Poveda et al.}, 1967; {\em
  Clarke and Pringle}, 1992; {\em Kroupa}, 2000).
The multiplicity of truly isolated field O-stars (if
they do exist, cf.  {\em de Wit et al.}, 2005; {\em Bressert et al.},
2012; {\em Oey et al.}, 2013) still needs to be investigated.

\subsection{Origin of Short-Period Massive Binary Systems} 
\label{sec:shortperiod}

In recent years it has become evident that at least 44\% of all O
stars are close spectroscopic binaries (see the review by {\em Sana
  and Evans}, 2011).  There are several -- at least five -- ideas to
explain the origin of such close massive spectroscopic binaries; these
are briefly discussed below.  In addition, we need to explain the
origin of hierarchical triple systems among massive stars; such
systems could either result from inner and outer disk fragmentation or
from a more chaotic dynamical N-body interaction.

Massive tight binaries cannot originate from the simple gravitational
fragmentation of massive cloud cores and filaments into two
Jeans-masses.  The Jeans-radius (10,000 AU) is far too large compared
with the separations of the two binary components (1-10 AU).  More
sophisticated physical processes must be at play, such as:

{\em (1) Inner disk fragmentation} ({\em Kratter and Matzner}, 2006)
followed by circumbinary accretion, to make the components grow in
mass ({\em Artymowicz and Lubow}, 1996).

{\em (2) Roche lobe overflow} of a close rapidly accreting bloated
proto-binary ({\em Krumholz and Thompson}, 2007).

In both cases, the authors argue that the accretion flow would drive
the component masses to near equality (massive twins). These theories,
however, do not explain how to get the initially lower-mass close
binaries in the first place.

{\em (3) Accretion onto a low-mass initially wide binary system} ({\em
  Bonnell and Bate}, 2005). While growing in mass by accretion, the
orbital separation of the binary system keeps shrinking.  In this
case, one can show analytically that -- depending on the angular
momentum of the accreting gas -- the wide binary, while growing in
mass, will shrink its orbital separation substantially (for example:
two 1~solar mass protostars at 30~AU separation can easily end up as
two 30~solar mass components at about 1~AU separation if the specific
angular momentum of the accreted gas is constant; that is, if
accreting gas angular momentum scales linearly with the accreted gas
mass).

{\em (4) Magnetic effects on fragmentation}. As noted in
Section~\ref{sec:physics}, 3D MHD calculations are just now becoming
commonplace, and effects such as magnetic torques on rotating clouds
might well lead to the formation of closer binary star systems 
than those found to date by 3D HD and RHD models of the fragmentation
process ({\em Price and Bate}, 2007).

{\em (5) Viscous evolution and orbital decay}. When a triple system
breaks up and ejects a component, the orbit of the remaining binary
tightens and becomes highly eccentric (see
Section~\ref{sec:dynamics}). When this occurs at early evolutionary
stages while the binary components are still surrounded by dense
circumstellar material, the components interact viscously during
periastron passages and their orbits decay (e.g., {\em Stahler}, 2010;
{\em Korntreff et al.}, 2012). At the same time, the UV radiation
field photoevaporates the circumstellar material, leaving many
binaries stranded in close orbits.

\subsection{Trapezia}
\label{sec:trapezia}

The famous Orion Trapezium (e.g., {\em Herbig and Terndrup}, 1986;
{\em Close et al.}, 2012) is the prototype of non-hierarchical
compact groups of OB stars. The concept was first introduced by {\em
  Ambartsumian} (1954), who recognized that such systems are
inherently unstable.  Kinematic studies of trapezia show the internal
motions expected for bound, virialized small clusters, but
occasionally having components with velocities exceeding the escape
speed ({\em Allen et al.}, 2004).

High precision astrometry from radio interferometry has demonstrated
that three of the sources in the Becklin-Neugebauer/Kleinman-Low
(BN/KL) region in Orion have large motions and are receding from a point in
between them, suggesting that they were all part of a small stellar
group, which disintegrated $\sim$500~yr ago ({\em Rodr\'\i guez et
  al.}, 2005; {\em G\'omez et al.}, 2006, but see {\em Tan}, 2004).
Just like the disintegration of small low-mass, very young stellar
systems can lead to giant Herbig-Haro bow shocks ({\em Reipurth},
2000), so will the break-up of a trapezium of massive protostars with
abundant gas lead to an energetic, explosive event, as observed around
the BN/KL region ({\em Bally and Zinnecker}, 2005; {\em Zapata et
  al.}, 2009; {\em Bally et al.}, 2011).

Trapezia are common in regions of massive star formation (e.g., {\em
  Salukvadze and Javakhishvili}, 1999; {\em Abt and Corbally}, 2000).
Of particular interest are studies of the earliest stages of formation
of a trapezium at centimeter and millimeter wavelengths (e.g., {\em
  Rod\'on et al.}, 2008). N-body simulations of massive trapezia in
clusters demonstrate that these systems are highly dynamical entities,
interacting and exchanging members with the surrounding cluster before
eventually breaking apart ({\em Allison and Goodwin}, 2011).

\section{\textbf{STATISTICAL PROPERTIES OF MULTIPLE STARS}}
\label{sec:statistics}

Multiple  systems  with  three  or  more  components  (hereafter  {\em
  multiples})  are a natural  and   rather  frequent  outcome  of  star
formation. Compared  to binaries, they have  more parameters (periods,
mass ratios,  etc.), so their statistics bring  additional insights on
the formation mechanisms. 

We focus here on stars with primary components of about one solar
mass, as their multiplicity statistics are known best.  {\em Raghavan
  et al.}  (2010) estimated a multiplicity fraction $MF = 0.46$ and a
higher-order fraction (triples and up) $HF \approx 0.12$ in a sample
of 454 solar-mass dwarfs within 25\,pc of the Sun.  A much larger
sample is needed, however, for a meaningful statistical study of
hierarchical systems.  Here we present preliminary results on F- and
G-dwarfs within 67\,pc selected from Hipparcos, the {\em FG-67pc}
sample ({\em Tokovinin}, 2014).  It contains a few hundred
hierarchical systems among $\sim$5000 stars.

\begin{figure}[ht]
 \epsscale{0.98}
\plotone{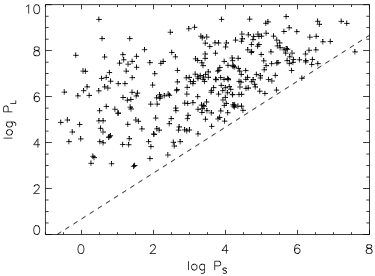}
  \caption{\small Orbital  periods $P_S$ at  inner hierarchical 
    levels~{\em 11} and {\em 12} are compared to  the periods 
    of outer systems $P_L$ from
    the FG-67pc sample.  The periods are expressed in days and plotted
    on  the logarithmic scale.   The dashed  line marks  the dynamical
    stability limit $P_L/P_S =4.7$.
\label{fig:plps}
}
 \end{figure}

 {\bf Period ratio and dynamical stability.}  Figure~\ref{fig:plps}
 compares the inner, {\em short} periods $P_S$ at levels~{\em 11} and
 {\em 12} to the outer, {\em long} periods $P_L$ at level~{\em 1} for the
 FG-67pc sample (for a definition of levels, see Section~3).  Note
 that orbital periods of wide pairs are estimated statistically by
 assuming that projected separation equals orbital semi-major axis.
 Such estimates $P^*$ are unbiased and differ from the true periods
 $P$ by less than 3 times, in most cases.

The  points in  Fig.~\ref{fig:plps} fill  the space  above  the dashed
line, reflecting  the fact  that all combinations  of inner  and outer
periods allowed dynamically are actually possible.  The minimum period
(or separation) ratio allowed  by dynamical stability has been studied
by several  authors.  The stability  criterion of Mardling  \& Aarseth
(2001), for example, can be written as
\begin{equation}
P_L/P_S > 4.7 (1 - e_L)^{-1.8}  (1 + e_L)^{0.6}  (1 + q_{\rm  out})^{0.1} ,
\end{equation}
where $e_L$ is the eccentricity of the outer orbit, while the ratio of
the distant-companion  mass to the  combined mass of the  inner binary
$q_{\rm  out}$  plays   only  a  minor  role.   The   dashed  line  in
Fig.~\ref{fig:plps}  corresponds  to $P_L/P_S  =4.7$;  all points  are
above  it (with  one exception  caused by  the uncertainty  of $P^*$).
Although orbits  of outer  systems tend to  have moderate  $e_L$ ({\em
  Shatsky},  2001),  its variation  spreads  the  value of the 
$P_L/P_S$ threshold over at least one order of magnitude.

Outer systems with $P<10^3$\,d do not exist or are rare (see the empty
lower-left corner in Fig.~\ref{fig:plps}). Such triples can be readily
discovered  by  radial-velocity  variations  superposed on  the  short
(inner)  orbit,  so  their  absence  is  not  an  observational  bias.
However,   tight  triples   are  found   among  massive   stars. 

{\bf   Distribution  of   periods   and  mass   ratios,  fraction   of
  hierarchies.}   Accounting   for  the  observational   selection  is
critical.   In the  FG-67pc  sample, the  probability  of detecting  a
companion to the  main target over the full range  of periods and mass
ratios  has been determined to be about  78\%.  This  means that  
only $0.78^2  \approx 0.61$
fraction  of level-{\em 11}  triples is  actually discovered  (assuming that
detections  of  inner and  outer  companions  are uncorrelated).   The
observed  fractions of  S:B:T:Q  systems are  64:29:6:1 percent.   The
selection-corrected fractions  are 54:31:6:7.  The  difference between
observed  (raw)  and  corrected  fractions increases  with  increasing
multiplicity.  Some  systems known presently  as binaries are  in fact
triples, some triples are quadruples, etc.

The joint distribution of period $P$ and mass ratios $q = M_2/M_1$ is
frequently  approximated  by  the  Gaussian  distribution of   $x  =
\log_{10} (P/1d)$ and by the power-law distribution of $q$  (see e.g., {\em
  Duch\^ene \& Kraus}, 2013):
\begin{equation}
f(x,q) = C\; \epsilon \; q^\beta \exp [ - (x -
    x_0)^2/(2 \sigma^2)] ,
\label{eq:Gauss}
\end{equation}
where  $\epsilon$  is   the  fraction  of  systems  and   $C$  is  the
normalization constant. It is  likely that the mass-ratio distribution
depends on period, but this is still being debated.  

The parameters of equation (\ref{eq:Gauss}) for the FG-67pc sample are
found by maximum likelihood, accounting for the incomplete detections
and missing data.  When all stellar pairs are considered regardless of
their hierarchical levels, the result is $\epsilon = CSF =0.57 \pm
0.02 $, while the median period is $x_0 = 4.53 \pm 0.09$.  If, on the
other hand, we count only the outer level-{\em 1} systems, the result
is different: $\epsilon = MF =0.47 \pm 0.01 $ and $x_0 = 4.97 \pm
0.06$.  Binary periods at the outer hierarchical level are thus almost
3 times longer than the periods of {\em all} binaries.  Similarly, for
the {\em inner} pairs at levels~{\em 11} and {\em 12} we derive much
shorter median periods $x_0=3.12$ and $x_0=2.45$, respectively. Note
that the formal errors quoted above are only lower statistical limits;
the results are influenced by several assumptions and approximations
made in the analysis, making the real uncertainty larger.  The
exponent of the mass-ratio distribution turns out to be small, $\beta
\approx 0.2$, meaning that the distribution of $q$ is almost uniform.

We derive the selection-corrected fractions of sub-systems of level~{\em 11}
and level~{\em 12} as 10\% and 8\%, respectively.  Discovery of sub-systems
in the {\em secondary} companions (level~{\em 12}) is more problematic than
at level~{\em 11}. Usually researchers concentrate on discovering companions
to their primary targets and forget that some of those companions may,
in turn, be close pairs.  The estimated detection rate of level-{\em 12}
sub-systems in the FG-67pc sample is only $\sim0.2$, so their true
frequency depends on the large, hence uncertain, correction. However,
there is a strong evidence that the occurrence of sub-systems in the
secondary components is nearly as frequent as in the main (primary)
targets.

Among the 88 sub-systems of level~{\em 12}, about a half also have
sub-systems of level~{\em 11}.  There is hence a correlation between
those levels: the frequency of 2+2 quadruples is larger than could be
inferred from the frequency of levels {\em 11} and {\em 12} if they
were independent (uncorrelated).  Among the 8\% of systems containing
secondary pairs of level~{\em 12}, half also contain level-{\em 11} pairs, they
are 2+2 quadruples.  Considering this, the fraction of systems with at
least 3 companions is $HF \approx 0.10 + 0.04 = 0.14$, not $0.10 +
0.08 = 0.18$ as one might naively assume by summing up the frequencies
of levels {\em 11} and {\em 12}.

{\bf   Statistical  model  of   hierarchical  multiplicity.}    It  is
remarkable   that   inner   pairs   in  hierarchical   multiples   are
statistically  similar  to  simple   binaries.   The  mass  ratios  in
spectroscopic  binaries with and  without distant  tertiary companions
are distributed  in the same way  ({\em Tokovinin et  al.}, 2006). The
frequency of  spectroscopic sub-systems in visual  binaries is similar
to  the frequency of  spectroscopic binaries  in the  open-cluster and
field populations  ({\em Tokovinin \& Smekhov,}  2002).  The frequency
of  resolved  sub-systems in  wide  binaries  is  again comparable  to
binaries in the field ({\em Tokovinin et al.,} 2010).  To first order,
we  can construct  a  hierarchical triple  by  selecting two  binaries
randomly from  the same {\em  generating distribution} of  periods and
keeping  only  stable  (hierarchical)  combinations.  This  recipe  is
applied recursively to simulate higher-order multiples.

To test this idea, {\em Tokovinin} (2014) simulated multiples,
filtered them by the average detection probability, and compared to
the real sample, following the strategy of {\em Eggleton} (2009). The
parameters of the generating distribution (equation \ref{eq:Gauss})
were taken from the maximum-likelihood analysis and could be further
adjusted to improve the agreement between the simulated and real
samples.  If the multiplicity fraction $\epsilon$ is kept constant,
the HF in the simulated sample is too low.  So, to reach an agreement
between simulations and reality, we had to increase $\epsilon$ at
inner hierarchical levels and to introduce a correlation between
levels~{\em 11} and {\em 12}.  Alternatively, the agreement can be
obtained by assuming a variable (stochastic) binary frequency
$\epsilon$.  Cases with a high $\epsilon$ produce many hierarchies,
while the cases of small $\epsilon$ generate mostly single and binary
stars.  This finding suggests that the field population is a mixture
coming from binary-rich and binary-poor environments.  Differences of
the multiplicity fraction among star-forming regions are well
documented (see Section 10.3).

Interestingly, the simulated  quadruples outnumber triples, resembling
in this respect the hydrodynamical simulations of {\em Bate} (2012). The
2+2 quadruples are much more  frequent ($\sim$4-5\% of all stars) than
the 3+1 quadruples. The large  number of 2+2 quadruples in the FG-67pc
sample predicted by this model can be verified observationally.

A loose correlation between orientations of the angular momentum
vectors in the inner and outer subsystems of triples was found in
early works and confirmed by {\em Tokovinin \& Sterzik} (2002).  This
correlation becomes stronger at moderate $P_L/P_S$ ratios, i.e.,  in
triples with weak hierarchy.  These authors tried to match the
observational result with simulations of dynamically decaying $N$-body
systems. Agreement could be achieved for certain initial conditions
(rotating and/or flattened clusters).  However, multiple systems
produced by the pure $N$-body decay without gas drag and accretion are
statistically very different from the real multiples in their
eccentricities and period ratios ({\em Tokovinin,} 2008), pointing to
the importance of viscous interactions and accretion during the
earliest phases of multiple evolution.

\section{\textbf{CONCLUDING REMARKS}}
\label{sec:conclusion}

In summary, it appears that the large majority -- and potentially all
-- of stars are born in small multiple systems.  A picture is emerging
where the field population of single, binary, and multiple stars
derives from a birth population that has been transformed by both
internal and external dynamical processes. These processes sculpt the
(still unknown) separation distribution function at birth into the
log-normal distribution (with a power-law tail for the wider binaries)
observed in the evolved field population.

\vspace{0.3cm}

\textbf{ Acknowledgments.}  

We thank the referee for a helpful report.
BR acknowledges support through the NASA Astrobiology Institute under
Cooperative Agreement No. NNA09DA77A issued through the Office of
Space Science.
APB's work was partially supported by the
NSF under grant AST-1006305.

\vspace{0.3cm}

\centerline{\textbf{ REFERENCES}}
\vspace{0.1cm}
\parskip=0pt
{\small
\baselineskip=11pt

\refs Abt, H. A. \& Corbally, C. J. (2000) {\em Astrophys. J., 541},
841.

\refs Albrecht, S. et al. (2009) {\it Nature
  461},373. 

\refs Albrecht, S. et al. (2011) {\it
  Astrophys. J. 726},68.

\refs Alexander, R. (2012) {\em Astrophys. J. Lett. 757}, L29.

\refs Allen, P. R. (2007) {\em Astrophys. J., 668}, 492.

\refs Allen, C. et al. (2004) IAU
Coll. No.191, eds. C. Allen \& C. Scarfe, {\em Rev. Mex. Astron.
  Astrofis. SC, 21}, 195.

\refs Allison, R. J. \& Goodwin, S. P. (2011)
{\em MNRAS, 415}, 1967.

\refs Ambartsumian, V. A. (1954) {\em Contrib. Byurakan Obs., 15}, 3.

\refs Anosova, J. P. (1986) {\em Astrophys. Spa. Sci., 124}, 217.

\refs Armitage, P. J. et al. (1999) {\em MNRAS, 304}, 425.

\refs Arreaga-Garc\'ia, G. et al. (2010) {\em Astron. Astrophys., 509}:A96.

\refs Artymowicz, P. (1983) {\em Acta. Astron., 33}, 223.

\refs Artymowicz, P. \& Lubow, S. H. (1994) {\em Astrophys. J., 421}, 651.

\refs Artymowicz, P. \& Lubow, S. H. (1996) {\em Astrophys. J., 467}, L77.

\refs Attwood, R. E. et al. (2009) {\em Astron. Astrophys., 495}, 201.

\refs Bai, X.-N. (2011) {\em Astrophys. J., 739}, 50.

\refs Baines, D. et al. (2006)
{\em MNRAS, 367}, 737.

\refs Bally, J. \& Zinnecker, H. (2005) {\em Astron. J., 129}, 2281.

\refs Bally, J. et al. (2011)
{\em Astrophys. J., 727}:A113.

\refs Baraffe I., \& Chabrier G. (2010) 
{\em Astron. Astrophys., 521}:A44.

\refs Baraffe I. et al. (1998)
{\em Astron. Astrophys., 337}, 403.

\refs Basu, S. \& Vorobyov, E.I. (2012) {\em Astrophys. J., 750}:A30.

\refs Bate, M. R. (2000) {\em MNRAS, 314}, 33.

\refs Bate, M. R. (2009a) {\em MNRAS, 392}, 590.

\refs Bate, M. R. (2009b) {\em MNRAS, 397}, 232.

\refs Bate, M. R. (2012) {\em MNRAS, 419}, 3115.

\refs Bate, M. R. \& Bonnell, I. A. (1997) {\em MNRAS, 285}, 33.

\refs Bate, M. R. et al. (2002) {\em MNRAS, 336}, 705.

\refs  Batten, A.  H.  (1973)  Binary and  Multiple  Star Systems,  {\em
  Pergamon Press, Oxford}

\refs Bechter, E. B., Creep, J. R., Ngo, H. et al. (2013) arXiv:1307.6857

\refs Becker, C. et al. (2013) {\em Astron. Astrophys., 552}:A46.

\refs Bergfors, C. et al. (2010) {\em Astron. Astrophys., 520}:A54.

\refs Biller, B. et al. (2011) 
{\em Astrophys. J., 730}:A39.

\refs Bisikalo, D. V. et al. (2012) {\em Astronomy Reports, 56}, 686.

\refs Blaauw, A. (1961) {\em Bull. Astr. Inst. Netherlands, 15}, 265.

\refs Boden, A. F. et al. (2009) {\em Astrophys. J 696}, L111.

\refs Bonnell, I. A. \& Bastien, P. (1992) {\em Astrophys. J., 401}, L31.

\refs Bonnell, I. A. \& Bate, M. R. (2005)
{\em MNRAS, 362}, 915.

\refs Bonnell, I. A. et al. (2007) in {\em
  Protostars and Planets V}, eds. B. Reipurth, D. Jewitt, K. Keil,
Univ. of Arizona Press, p. 149.

\refs Bonnell, I. A. et al. (2008) {\em MNRAS, 389}, 1556.

\refs Boss, A. P. (2009) {\em Astrophys. J., 697}, 1940.

\refs Boss, A. P. \& Bodenheimer (1979) {\em Astrophys. J., 234}, 289.

\refs Boss, A. P. \& Keiser, S. A. (2013) {\em Astrophys. J., 763}:A1.

\refs B\"urzle, F. et al. (2011) {\em MNRAS,  412}, 171.

\refs Brandeker, A. et al. (2006) {\em
  Astrophys. J., 652}, 1572.

\refs Bressert, E.  et al. (2012) {\em
  Astron. Astrophys., 542}, 49.

\refs Burgasser, A. J. et al. (2007) in {\em
  Protostars and Planets V}, eds. B. Reipurth, D. Jewitt, K. Keil,
Univ. of Arizona Press, 427.

\refs Cabrit S. et al. (2006)
{\em Astron. Astrophys., 452}, 897.

\refs Capelo, H. et al. (2012) {\em Astrophys. J., 757}:L18

\refs Carney, B. W. et al. (2003) {\em Astron. J., 125}, 293.

\refs Chabrier G. et al. (2007)
{\em Astron. Astrophys., 472}, L17.

\refs Chauvin, G. et al. (2011) {\it Astron. \& Astrophys. 528}:A8.

\refs Chen, X. et al. (2008) {\em Astrophys. J., 683}, 862.

\refs Chen, X. et al. (2009) {\em Astrophys. J., 691}, 1729.

\refs Chen, X. et al. (2013) {\em
  Astrophys.J., 768}:A110.

\refs Chini, R. et al. (2012) {\em MNRAS, 424}, 1925.

\refs Cieza, L. et al. (2009) {\em
  Astrophys. J., 696}, L84.

\refs Clarke, C. J. \& Pringle, J. E. (1992) {\em MNRAS, 255}, 423.

\refs Clarke, C. J. \& Pringle, J. E. (1993) {\em MNRAS, 261}, 190.

\refs Clark, P. C. et al. (2008) {\em MNRAS, 386}, 3.

\refs Close, L. M. et al. (2012)
{\em Astrophys. J., 749}:A180.

\refs Commercon, B. et al. (2010) {\em Astron. Astrophys., 510}:L3. 

\refs Commercon, B. et al. (2011) {\em Astrophys. J., 742}:L9. 

\refs Connelley, M. S. et al. (2008a) {\em
  Astron. J, 135}, 2496.

\refs Connelley, M. S. et al. (2008b) {\em
  Astron. J, 135}, 2526.

\refs Connelley, M. S. et al. (2009) {\em
  Astron. J, 138}, 1193.

\refs Corporon, P. \& Lagrange, A.-M. (1999) {\em Astron. Astrophys.
  Suppl., 136}, 429.

\refs Correia, S. et al. (2006)
{\em Astron. Astrophys., 459}, 909.

\refs Correia, S. et al. (2013)
{\em Astron. Astrophys., 557}:A63.

\refs Crutcher, R. M. (2012) {\em Ann. Rev. Astron. Astrophys., 50}, 29.

\refs Cuadra, J. et al. (2009) {\em MNRAS, 393}, 1423.

\refs Daemgen, S. et al. (2012) {\em
  Astron. Astrophys., 540}:A46.

\refs Daemgen, S. et al. (2013) {\em
  Astron. Astrophys., 554}:A43.

\refs D'Antona F. \& Mazzitelli I. (1997) {\em Mem. Soc. Astr. Ital., 68},
807.

\refs Deller, A.~T. et al. (2013) {\em Astron. Astrophys., 552}:A51. 

\refs De Rosa, R.J. et al. (2014) {\em MNRAS, 437}, 1216.

\refs de Val-Borro, M. et al. (2011) {\em MNRAS, 413}, 2679.

\refs de Wit, W. J. et al. (2005)
{\em Astron. Astrophys., 437}, 247.

\refs Domingos, R. et al. (2012) {\it Astron. Astrophys., 544}:A63. 

\refs Dotter A. et al.  (2008) {\em
  Astrophys. J. Supp., 178}, 89.

\refs Doyle, L. R. et al. (2011) {\em
  Science, 333}, 1602.

\refs Duch\^ene, G. \& Kraus, A. (2013) {\em Ann. Rev. Astron. Astrophys., 51}, 269 

\refs Duch\^ene, G. et al. (1999) {\em Astron.
  Astrophys., 343}, 831.

\refs Duch\^ene, G. et al.
(2004) {\em Astron. Astrophys., 427}, 651.

\refs Duch\^ene, G. et al. (2007) {\em Astron. Astrophys., 476}, 229.

\refs Dumusque, X. et al (2012) {\it Nature, 491}, 207.

\refs Dupuy, T. J. \& Liu, M. C. (2011) {\em Astrophys. J., 733}:A122.

\refs Duquennoy, A. \& Mayor, M. (1991) {\em Astron. Astrophys.,
  248}, 485.

\refs Durisen, R. H. \& Sterzik, M.F. (1994) {\em Astron. Astrophys.,
  286}, 84.

\refs Dyck, H. M. et al. (1982) {\em Astrophys. J.,
  255}, L103.

\refs Eggleton, P. (2009) {\em MNRAS, 399}, 1471.

\refs Enoch, M. L. et al. (2011) {\em Astrophys. J. Suppl. Ser., 195}:A21.

\refs Espaillat, C. et al. (2011) {\em
  Astrophys. J. 728}, 49.

\refs Fabrycky, D. \& Tremaine, S., (2007) {\it Astrophys. J., 669}, 1298.

\refs Facchini, S. \& Lodato, G., (2013) {\it MNRAS, 433}, 2142

\refs Fateeva, A. M. et al.
(2011) {\em Astrophy. Space Sci., 335}, 125.

\refs Feiden, G. \& Chaboyer, B. (2012) {\em Astrophys. J., 761}, 30.

\refs Feiden, G. A. \& Dotter, A. (2013) {\em Astrophys. J., 765}:A86.

\refs Fischer, D. A. \& Marcy, G. W. (1992) {\em Astrophys. J., 396}, 178.

\refs Foucart, F. \& Lai, D., (2013) {\em Astrophys. J., 764}:A106.
   
\refs Garcia, V. et al. (2011) {\em Astron. J.,
  142}, 27.

\refs Garcia, V. et al. (2013a)
{\em Astrophys. J., 769}, 114.

\refs Garcia, P. J. V. et al. (2013b) {\em MNRAS, 430}, 1839.

\refs Ghez, A. M. et al. (1993) {\em Astron.
  J., 106}, 2005.

\refs Ghez, A. M. et al. (1997)
{\em Astrophys. J., 481}, 378.

\refs Gieles, M. \& Portegies Zwart, S. F. (2011) {\em MNRAS, 410}, L6.

\refs Gies, D.R. \& Bolton, C.T. (1986) {\em Astrophys. J. Suppl.,
  61}, 419.

\refs Gomez da Castro et al.
(2013) {\em Astrophys. J., 766}, 62.

\refs G\'omez, L. et al. (2006)
{\em Astrophys. J., 635}, 1166.

\refs G\'omez Maqueo Chew, Y. et al. (2012)
{\em Astrophys. J., 745}, 58.

\refs Goodman, J. \& Hut, P. (1993) \apj, {\em 403}, 271.

\refs Goodwin, S. P. (2010) {\em Phil.Trans.Roy.Soc.A}, {\em 368}, 851.

\refs Goodwin, S. P. et al. (2007) in {\em Protostars and Planets V}
eds. B. Reipurth, D. Jewitt, K. Keil,  Univ. of Arizona Press, p. 133.

\refs Gorti, U. \& Bhatt, H. C. (1996) {\em MNRAS, 283}, 566.

\refs Haisch, K. E. et al. (2004) {\em
  Astron. J., 127}, 1747.

\refs Hall, S. (1997) {\it MNRAS, 287}, 148. 

\refs Hall, S. et al. (1996) {\em MNRAS, 278}, 303.

\refs Hallinan, G. et al. (2007) {\em
  Astrophys. J., 663}, L25.

\refs Hanawa, T. et al. (2010) {\em Astrophys. J., 708}, 485.

\refs Hansen, C. et al. (2012) {\it Astrophys.
  J., 747}:A22. 

\refs Harris, R. J. et al. (2012) {\em Astrophys. J., 751}:A115.  

\refs Heggie, D. C. (1975) {\em MNRAS, 173}, 729.

\refs Hennebelle, P. \& Fromang, S. (2008) {\em Astron. Astrophys.,
  477}, 9.

\refs Hennebelle, P. \& Teyssier, R. (2008) {\em Astron. Astrophys.,
  477}, 25.

\refs Hennebelle, P. et al. (2011) {\em Astron. Astrophys., 528}:A72.

\refs Herbig, G. H. (1962) {\em Advances Astron. Astrophys., 1}, 47.

\refs Herbig, G. H. (1977) {\em Astrophys. J., 217}, 693.

\refs Herbig, G. H. \& Terndrup, D. M. (1986) {\em Astrophys. J., 307}, 609.

\refs Herbst, W. et al. (2010) {\em Astron.
  J., 140}, 2025.

\refs Hills, J. G. (1975) \aj, {\em 80}, 809.

\refs Hills, J. G. (1990) \aj, {\em 99}, 979.

\refs Hioki, T. et al. (2009) {\em Pub. Astr. Soc. Japan, 61}, 1271.

\refs Hioki, T. et al. (2011) {\em Pub. Astr. Soc. Japan, 63}, 543.

\refs Jensen, E. L. et al. (1996) {\em
  Astrophys. J., 458}, 312.

\refs Harris, R. J. at al. (2012) {\em Astrophys. J., 751}:A115.

\refs Joergens, V. (2008) {\em Astron. Astrophys., 492}, 545.

\refs Joos, M. et al. (2012) {\em Astron. Astrophys., 543}:A128.

\refs Joy, A. H. \& van Biesbroeck, G. (1944) {\em PASP, 56}, 123.

\refs Kennedy, G. M.  et al. (2012a)
{\it MNRAS, 421}, 2264.

\refs Kennedy, G. M. et al. (2012b)
{\em MNRAS, 426}, 2115. 

\refs Kiminki, D.C. \& Kobulnicky, H.A. (2012) {\em Astrophys. J.,
  751}, 4.

\refs King, R. R. et al.
(2012a) {\em MNRAS, 421}, 2025.

\refs King, R. R. et al. (2012b) 
  {\em MNRAS, 427}, 2636.

\refs Kneller, S. A. \& Clarke, C. J. (2014) in prep.

\refs K\"ohler, R. et al.
  (2006) {\em Astron. Astrophys.}, {\em 458}, 461.

\refs Korntreff, C. et al. (2012) 
{\em Astron. Astrophys., 543}:A126.

\refs Kouwenhoven, M. B. N. et al.
(2005) {\em Astron. Astrophys., 430}, 137.

\refs Kouwenhoven, M. B. N. et al. (2007) {\em Astron. Astrophys., 474}, 77.

\refs Kouwenhoven, M. B. N. et al. (2010) {\em MNRAS, 404}, 1835.

\refs Kozai, Y. (1962) {\em Astron. J., 67}, 591.

\refs Kratter, K.M. \& Matzner, C. D. (2006) {\em MNRAS, 373}, 1563.

\refs Kraus, A. L. \& Hillenbrand, L. A. (2007) {\em Astrophys. J., 662}, 413.

\refs Kraus, A. L. \& Hillenbrand, L. A. (2009) {\em Astrophys. J., 784}, 531.

\refs Kraus, A. L. \& Hillenbrand, L. A. (2012) {\em Astrophys. J., 757}:A141.

\refs Kraus, A. L. et al.
(2008) {\em Astrophys. J., 679}, 762.

\refs Kraus, A. L. et al.
(2011) {\em Astrophys. J. 731}:A8.

\refs Kraus, A. L. et al. (2012) {\em Astrophys. J., 745}:A19.

\refs Kroupa, P. (1995) {\em MNRAS, 277}, 1491.

\refs Kroupa, P. (1998) {\em MNRAS, 298}, 231.

\refs Kroupa, P. (2000)
in {\em Massive Stellar Clusters}, (A. Lancon and C. Boily, ed.)
{\em ASP Conference Series, 211}, 233.

\refs Kroupa, P. \& Bouvier, J. (2003) {\em MNRAS, 346}, 343.

\refs Kroupa, P. \& Petr-Gotzens, M. G. (2011) {\em Astron.
  Astrophys., 529}:A92.

\refs Kroupa, P. et al. (2003)
{\em MNRAS, 346}, 354.

\refs Krumholz, M. R. \& Thompson, T. A. (2007) {\em Astrophys. J.,
  661}, 1034.

\refs Krumholz, M. et al. (2012) {\it Astrophys. J., 754}:A71. 

\refs Kudoh, T. \& Basu, S. (2008) {\em Astrophys. J., 679}, L97.

\refs Kudoh, T. \& Basu, S. (2011) {\em Astrophys. J., 728}:A123.

\refs Kuiper, G. P. (1942) {\em Astrophys. J., 95}, 201.

\refs Lada, C. J. (2006) {\em Astrophys. J., 640}, L63.

\refs Lafreni\`ere, D. et al. (2008)
{\em Astrophys. J., 683}, 844.

\refs Larson, R. B. (1972) {\em MNRAS, 156}, 437.

\refs Lazarian, A. et al. (2012) {\em Astrophys.
  J., 757}:A154.

\refs Leigh, N. \& Geller, A. M. (2012) {\em MNRAS, 425}, 2369.

\refs Leinert, Ch. et al. (1993) {\em Astron. Astrophys., 278}, 129.

\refs Leinert, Ch. et al. (1997) {\em Astron.
  Astrophys., 318}, 472.

\refs Leinhardt, Z. \& Stewart, S. (2012) {\it Astrophys. J., 745}, 79.

\refs L\'epine, S. \& Bongiorno, B. (2007) {\em Astron. J., 133}, 889.

\refs Lidov, M., (1962) {\em Plan. Spa. Sci., 9}, 719.

\refs Loinard, L.\ (2013) IAU Symposium 289, {\em Advancing the
  Physics of Cosmic Distances}, ed. R. de Grijs, 36.

\refs Looney, L. G. et al. (2000) {\em Astrophys.
  J., 529}, 477.

\refs MacDonald J. \& Mullan D.J. (2009)
{\em Astrophys. J., 700}, 387.

\refs MacFadyen, A. I. \& Milosavljevic, M. (2008) {\em Astrophys.
  J., 672}, 83.

\refs Machida, M. N. (2008) {\em Astrophys. J., 682}, L1.

\refs Machida, M. N. et al. (2008) {\em Astrophys. J., 677}, 327.

\refs Maiz-Appellaniz, J. et al. (2004)
{\em Astrophys. J. Suppl. 151}, 103.

\refs Mardling, R. A. \& Aarseth, S. J. (2001) {\em MNRAS, 321}, 398.

\refs Marks, M. \& Kroupa, P. (2011) {\em MNRAS, 417}, 1702.

\refs Marks, M. \& Kroupa, P. (2012) {\em Astron. Astrophys., 543}:A8.

\refs Martins, F. et al. (2005) {\em Astron. Astrophys., 436}, 1049.

\refs Marzari, F. et al. (2013) {\em Astron. Astrophys., 553}:A71.

\refs Mason, B. D. et al.  (2009) {\em
  Astron. J., 137}, 3358.

\refs Mathieu, R. D. (1994) {\em Ann. Rev. Astron. Astrophys., 32}, 465.

\refs Mathieu, R. D. et al. (1997) {\em Astron. J., 113}, 1841.

\refs Mathieu R. D. et al.
(2007) in {\em Protostars \& Planets V}, eds. B. Reipurth, D. Jewitt, K. Keil,
Univ. of Arizona Press, p. 411.

\refs Maury, A. J. et al. (2010) {\em Astron. Astrophys., 512}:A40.

\refs Melo, C. H. F. et al. (2001)
{\em Astron. Astrophys., 378}, 898.

\refs Moeckel, N. \& Bate, M. R. (2010) {\em MNRAS, 404}, 721.

\refs Moeckel, N. \& Clarke, C. J. (2011) {\em MNRAS, 415}, 1179.

\refs Mohanty, S. \& Stassun, K. G. (2012) {\em Astrophys. J., 758}, 12. 

\refs Mohanty, S. et al. (2010)
{\em Astrophys. J., 722}, 1138.

\refs Mohanty, S. et al. (2013) {\em
  Astrophys. J., 764}, 65.

\refs Monin, J.-L. et al (2007) in {\em
  Protostars and Planets V},  eds. B. Reipurth, D. Jewitt, K. Keil,
Univ. of Arizona Press, p. 395.

\refs Morales, J.C. et al. (2010) {\em
  Astrophys. J., 718}, 502.

\refs Morales-Calderon, M. et al. (2012)
{\em Astrophys. J., 753}, 149.

\refs Muzerolle, J. et al. (2013) {\em
  Nature, 493}, 378.

\refs Myers, A. et al. (2013) {\it
  Astrophys. J., 766}:A97.

\refs Nagel, E. et al. (2011) {\em
  Astrophys. J., 747}, 139.

\refs Nguyen, D. C. et al.
(2012) {\em Astrophys. J., 745}:A119.

\refs Ochi, Y. et al. (2005) {\em Astrophys. J.,
  623}, 922.

\refs Oey, M.S.  et al. (2013) {\em
  Astrophys. J., 768}, 66.

\refs Offner, S. S. R. et al. (2009) {\em
  Astrophys. J., 703}, 131.

\refs Offner, S. S. R. et al. (2010) {\it
  Astrophys. J., 725}, 1485.

\refs \"Opik, E. (1924) {\em Pub. Tartu Obs., 25}, No.6.

\refs Padoan, P. \& Nordlund, \AA. (2004) {\em Astrophys. J., 617}, 559.

\refs Palla F. \& Stahler S. (1999) {\em Astrophys. J., 525}, 772.

\refs Parker, R. J. \& Goodwin, S. P. (2012) {\em MNRAS, 424}, 272.

\refs Parker, R. J. \& Reggiani, M. M. (2013) {\em MNRAS, 432}, 2378.

\refs Parker, R. J. et al. (2009)  {\em MNRAS, 397}, 1577.

\refs Parker, R. J. et al. (2011) {\em MNRAS, 418}, 2565.

\refs Payne, M. et al. (2009) {\it MNRAS, 336}, 973.

\refs Pepper, J. et al. (2007) {\it Pub. Astr.
  Soc. Pacific, 119}, 923.

\refs Pfalzner, S. et al. (2005) {\it Astrophys.
  J., 629}, 526.

\refs Poveda, A. et al. (1967)
{\em Bol. Observ. Ton. Tac., 4}, 86.

\refs Prato, L. (2007) {\em Astrophys. J., 657}, 338.

\refs Prato, L. \& Simon, M. (1997) {\em Astrophys. J., 474}, 455.

\refs Price, D. J. \& Bate, M. R. (2007) {\em MNRAS, 377}, 77.

\refs Rafikov, R. (2013) {\it Astrophys. J., 765}, L8.

\refs Raghavan, D. et al. (2010) {\em Astrophys. J. Suppl., 190}, 1.

\refs Ratzka, T. et al. (2005) {\em Astron.
  Astrophys., 437}, 611.

\refs Rawiraswattana, K. et al. (2012)
{\em MNRAS, 419}, 2025.

\refs Reid, I. N. et al. (2006) {\em
  Astron. J., 132}, 891.

\refs Reiners, A. et al. (2007)
{\em Astrophys. J., 671}, L149.

\refs Reipurth, B. (1988) in {\em Formation and Evolution of Low-Mass
  Stars}, eds. A.K. Dupree, M.T.V.T. Lago (Kluwer), p. 305.

\refs Reipurth, B. (2000) {\em Astron. J., 120}, 3177.

\refs Reipurth, B. \& Bally, J. (2001)
{\em Ann. Rev. Astron. Astrophys., 39}, 403.

\refs Reipurth, B. \& Clarke, C. (2001) {\em Astron. J., 122}, 432.

\refs Reipurth, B. \& Aspin, C. (2004)
{\em Astrophys. J., 608}, L65.

\refs Reipurth, B. \& Mikkola, S. (2012) {\em Nature, 492}, 221.

\refs Reipurth, B. \& Mikkola, S. (2014) in prep.

\refs Reipurth, B. \& Zinnecker, H. (1993) {\em Astron. Astrophys.,
  278}, 81.

\refs Reipurth, B. et al. (2002) {\em Astron. J., 124}, 1045.

\refs Reipurth, B. et al. (2004) {\em Astron. J., 127}, 1736.

\refs Reipurth, B. et al. (2007) \aj, {\em 134}, 2272.

\refs Reipurth, B. et al. (2010)
{\em Astrophys. J., 725}, L56.

\refs Rod\'on, J. A. et al. (2008) {\em Astron. Astrophys., 490}, 213.

\refs Rodriguez, J. et al. (2013) {\em
  Astron. J., 146}:A112. 

\refs Rodr{\'{\i}}guez, L.~F. et al. (2003)
{\em Astrophys. J., 598}, 1100.

\refs Rodr\'\i guez, L. F. et al. (2005)
{\em Astrophys. J., 627}, L65.

\refs Rodr{\'{\i}}guez, L.~F. et al. (2010) {\em Astron. J., 140}, 968. 

\refs Salter, D. M. et al. (2010) {\em
  Astron. Astrophys., 521}:A32.

\refs Salukvadze, G. N. \& Javakhishvili, G. Sh. (1999)
{\em Astrophysics, 42}, 431.

\refs Sana, H. \& Evans, C.J. (2011) 
{\em IAU Symp., 272}, 474.

\refs Sana, H. et al. (2008) 
{\em MNRAS, 386}, 447.

\refs Sana, H. et al. (2009) {\em MNRAS, 400}, 1479.

\refs Sana, H. et al. (2012) 
{\em Science 337}, 444.

\refs Sana, H. et al. (2013) 
{\em Astron. Astrophys., 550}, 107.

\refs Scally, A. et al. (1999) 
{\em MNRAS, 306}, 253.

\refs Scally, A. et al. (2005)
{\em MNRAS, 358}, 742.

\refs Seifried, D. et al.
(2012) {\em MNRAS, 423}, L40.

\refs Shakura, N. \& Sunyaev, R. A. (1973) {\em Astron. Astrophys.,
  24}, 337.

\refs Shatsky, N. (2001) {\em Astron. Astrophys., 380}, 238.

\refs Shi, J.-M. et al. (2012) {\em Astrophys.
  J., 747}, 118.

\refs Siess I. et al. (2000)
{\em Astron. Astrophys., 358}, 593.

\refs Simon M. \& Obbie R.C. (2009) {\em Astron. J., 137}, 3442.

\refs Sota, A. et al.
(2008) {\em Rev. Mex. Astron. Astrofis. S.C., 33}, 56. 

\refs Stahler, S. W. (2010) {\em MNRAS, 402}, 1758.

\refs Stamatellos, D. et al. (2007) {\em
  MNRAS, 382}, L30.

\refs Stassun, K.G. et al. (2006) {\em
  Nature, 440}, 311.

\refs Stassun, K.G. et al. (2007)
{\em Astrophys. J., 664}, 1154.

\refs Stassun, K.G. et al.
 (2008) {\em Nature, 453}, 1079.

\refs Stassun, K.G. et al. (2012)
{\em Astrophys. J., 756}, 47.

\refs Stempels, H. \& Gahm, G. (2004) {\em Astron. Astrophys.
   421}, 1159.

 \refs Sterzik, M.  \& Durisen, R.  (1998) {\em Astron.
   Astrophys., 339}, 95.

 \refs Sterzik, M.  \& Tokovinin, A. (2002) {\em Astron.
   Astrophys., 384}, 1030.

 \refs Sterzik, M.  et al. (2003) {\em
   Astron. Astrophys., 411}, 91.

 \refs Tan, J. (2004) {\em Astrophys. J., 607}, L47.

\refs Thebault, P. et al. (2006) {\it Icarus 183}, 193.

\refs Thebault, P. et al. (2008) {\it MNRAS, 388}, 1528.

\refs Thebault, P. et al. (2009) {\it MNRAS, 393}, L21.

\refs Tobin, J. J. et al. (2009) {\em Astrophys. J., 697}, 1103.

\refs Tobin, J. J. et al. (2013) {\em Astrophys. J., 779}:A93.

\refs Tognelli E. et al. (2011)
{\em Astron. Astrophys., 533}, 109.

\refs Tokovinin, A. \& Smekhov, M. (2002) {\em Astron.
  Astrophys., 382}, 118.

\refs Tokovinin, A. et al. (2006) {\em
  Astron. Astrophys., 450}, 681.

\refs Tokovinin, A. (2008)  {\em MNRAS, 389}, 925.

\refs Tokovinin, A. (2014), in preparation.

\refs Tokovinin, A. \& L\'epine, S. (2012) {\em Astron. J., 144}:A102.

\refs Tokovinin, A. et al. (2010) {\em
  Astron. J., 140}, 510.

\refs Torres, G. (2013) {\em Astron. Nach., 334}, 4.

\refs Torres, G. et al. (2010)
{\em Astron. Astrophys. Rev., 18}, 67.

\refs Torres, G. et al.
(2013) {\em Astrophys. J., 773}:A40.

\refs Torres, R.~M. et al. (2012)
{\em Astrophys. J., 747}, 18.

\refs Trilling, D. et al. (2007)
{\em Astrophys. J., 658}, 1264.

\refs Umbreit, S. et al. (2005) {\em
  Astrophys. J., 623}, 940.

\refs Umbreit, S. et al. (2011) {\em
  Astrophys. J., 743}:A106.

\refs Valtonen, M. \& Karttunen, H. (2006) {\em The Three-Body
  Problem}, Cambridge Univ. Press.

\refs van Albada, T. S. (1968) {\em Bull. Astron. Inst. Netherlands, 20}, 57.

\refs Verrier, P. \& Evans, N. (2007) {\it MNRAS, 382}, 1432.

\refs Verrier, P. \& Evans, N. (2009) {\it MNRAS, 394}, 1721.

\refs Viana Almeida, P. et al. (2012) {\em Astron.
  Astrophys., 539}:A62.

\refs Walch, S. et al. (2010) {\em MNRAS, 402}, 2253.

\refs Welsh, W. F. et al. (2012) {\em
  Nature, 481}, 475.

\refs Whitworth, A. P. \& Zinnecker, H. (2004) {\em Astron.
  Astrophys., 427}, 299.

\refs Whitworth, A. P. et al. (2007) in {\em Protostars and Planets V}, eds. B. Reipurth, D. Jewitt, K. Keil,  Univ. of Arizona Press, p. 459.

\refs Zhao, B. \& Li,  Z.-Y. (2013) {\it Astrophys. J., 763}, 7.

\refs Xie, J.-W. et al. (2010) {\it Astrophys. J.,
  708}, 1566.

\refs Zahn, J.-P. \& Bouchet, L. (1989) {\em Astron. Astrophys., 223}, 112.

\refs Zapata, L. A. et al. (2009)
{\em Astrophys. J., 704}, L45.

\refs Zinnecker, H. (1989) in ESO Workshop {\em Low Mass Star
  Formation and Pre-Main Sequence Objects}, ed. Bo Reipurth, p. 447.

\refs Zinnecker, H. \& Mathieu, R. (2001) {\em The Formation of Binary
  Stars}, IAU Symp. No. 200, Astron. Soc. Pacific.

\refs Zinnecker, H. \& Yorke, H.W. (2007) 
{\em Ann. Rev. Astron. Astrophys., 45}, 481.

\end{document}